\documentclass[12pt]{article}
\pdfoutput=1

\usepackage{amsmath,amsfonts,amssymb,epsfig}
\usepackage{slashed}
\usepackage[width=0.9\textwidth,margin={30pt,30pt},font=small,labelfont=bf]{caption}
\numberwithin{equation}{section}

\usepackage{hyperref}
\usepackage{enumerate}

\usepackage{bbold}
\usepackage{xcolor}
\hypersetup{
    colorlinks,
    linkcolor={red!0!black},
    citecolor={blue!50!black},
    urlcolor={blue!80!black}
}
\usepackage{array}
\usepackage{subfig}
\usepackage{subcaption}
\usepackage{graphicx}
\usepackage{cancel}
\usepackage{color}
\usepackage{bm}
\usepackage{multirow}
\usepackage{multicol}
\usepackage{xspace}
\usepackage{tcolorbox}
\usepackage{adjustbox}
\usepackage{tikz}

\usepackage{tikz-feynman}
\tikzfeynmanset{compat=1.0.0}

\usepackage{cite}

\newcommand{\exclude}[1]{}
\newcommand{\al}{\alpha}

\newcommand{\psib}{\bar{\psi}}

\newcommand{\bone}{\boldsymbol{1}}
\newcommand{\btwo}{\boldsymbol{2}}
\newcommand{\bthree}{\boldsymbol{3}}

\newcommand{\bp}{\boldsymbol{p}}

\newcommand{\dotalpha}{\Dot{\alpha}}
\newcommand{\dotbeta}{\Dot{\beta}}

\newcommand{\TG}[1]{{\color{blue} TG: #1}}
\newcommand{\WK}[1]{{\color{red} WK: #1}}

\begin{document}

\hypersetup{pageanchor=false}
\begin{titlepage}

\begin{center}

\hfill UMN-TH-4332/24 \\
\hfill FTPI-MINN-24-19\\

\vskip 0.5in

{\Large \bfseries Effective interactions and on-shell\vspace{2mm} \\  recursion relation for massive spin 3/2} \\
\vskip .8in

{\small\bf Tony Gherghetta,$^{1,a}$} \let\thefootnote\relax\footnote{$^a$tgher@umn.edu}
{\small\bf Wenqi Ke$^{1,2,b}$}
\footnote{$^b$wke@umn.edu}
\begin{tabular}{ll}
$^{1}$ & \!\!\!\!\!\emph{\footnotesize School of Physics and Astronomy, University of Minnesota, Minneapolis, Minnesota 55455, USA}\\
$^{2}$ & \!\!\!\!\!\emph{\footnotesize  William I. Fine Theoretical Physics Institute, School of Physics and Astronomy,}\\[-.15em]
& \!\!\!\!\!\emph{\footnotesize  University of Minnesota, Minneapolis, Minnesota 55455, USA}\\
\end{tabular}

\end{center}
\vskip .6in

\begin{abstract}
\noindent
We use on-shell methods to compute all three-point interactions of massive spin-3/2 particles involving a graviton and particles of spin $\leq 1$. By employing the massive spinor-helicity formalism we identify the interactions which have a smooth massless limit as expected from the superHiggs mechanism. These interactions are then used to on-shell construct four-point massive spin-3/2 amplitudes using an all-line transverse shift for the external momenta, which correctly reproduces the contact gravitino interactions in the $N=1$ supergravity Lagrangian. The on-shell constructed four-point amplitudes are also used to derive well-known unitarity bounds in supergravity. In particular, by adding scalar and pseudoscalar interactions to construct the four-point massive spin-3/2 amplitudes that scale as $E^2$ in the high-energy limit, we recover the on-shell Polonyi model with a Planck scale unitarity bound. These effective three-point interactions and on-shell recursion relations provide an alternative and simpler way to study the interactions of massive spin-3/2 particles without a Lagrangian or the use of Feynman diagrams. 
\end{abstract}

\end{titlepage}

\tableofcontents

\newpage
 \section{Introduction}

On-shell methods have provided remarkable new insight and ease into the calculation of tree-level scattering amplitudes.  
These methods rely on the spinor-helicity formalism (see~\cite{Dixon:2013uaa,Elvang_Huang_2015}, for a review) which allows for the construction of effective three-point interactions using little-group covariance. 
These interactions can then act as the building blocks to construct higher-point amplitudes using on-shell recursion relations.
This method first developed for gauge theory and gravity~\cite{Britto:2004ap,Britto:2005fq} has been subsequently extended to other massless theories~\cite{Arkani-Hamed:2008owk,Cohen:2010mi,Cheung:2015ota} as well as massive theories~\cite{Badger:2005zh,Badger:2005jv,Ozeren:2006ft,Ballav:2020ese}. The main idea is to analytically continue an amplitude into the complex plane by shifting external momenta parametrized by a complex variable $z$. The complex poles of the amplitude then correspond to intermediate particles going on-shell where the amplitude factorizes into lower-point amplitudes. By using Cauchy's theorem from complex analysis, the physical amplitude at $z=0$ can then be obtained by contour integration provided the contribution from the boundary at $|z|\rightarrow \infty$ vanishes. This provides an efficient method to compute scattering amplitudes without recourse to Lagrangians or Feynman diagrams.

In this work we extend these on-shell methods to include spin-3/2 particles that are Majorana. As is well-known, spin-3/2 particles play an important role in locally supersymmetric (or supergravity) theories, suggesting that they are likely to exist in any UV completion involving gravity. Moreover, their mass is related to the scale of supersymmetry breaking which provides motivation to study massive spin-3/2 particles. In general, since spin-3/2 particles arise from a tensor product of spin-1 and spin-1/2 Lorentz group representations, they inherit redundant features associated with gauge bosons and fermions, 
which typically makes the study of spin-3/2 interactions using off-shell Lagrangians and Feynman diagrams cumbersome. Therefore, on-shell methods are ideally suited to simplify the classification and constructibility of amplitudes involving spin-3/2 particles.

Using the massive spinor-helicity formalism developed in~\cite{Arkani-Hamed:2017jhn}, we first catalog all massive spin-3/2 interactions involving the graviton and particles of spin $\leq 1$ and verify well-known results. In particular, we find that the three-point vertex involving two massive spin-3/2 particles and a gauge boson is purely vector-like and vanishes due to the antisymmetry of the Majorana particles.
Instead, for the coupling of two spin-3/2 particles to the graviton, the first term in a power series expansion of the amplitude coincides with the gravitational minimal coupling obtained from the energy-momentum tensor. We also derive various three-point vertices involving just a single spin-3/2 particle. This enables one 
to determine the spin-3/2 lifetime without any reference to a Lagrangian, relying only on the number of parity even and odd interactions. Furthermore, in the high-energy (or massless) limit, we verify that the massive spin-3/2 vertices correctly reduce to the three-point interactions involving a massless spin-1/2 fermion (associated with the spin-3/2 longitudinal component) as expected by the equivalence theorem~\cite{Casalbuoni:1988kv,Casalbuoni:1988qd} or as advocated in Ref.~\cite{Arkani-Hamed:2017jhn} due to an ``IR unification" manifesting as the superHiggs effect.

Armed with the effective three-point interactions involving massive spin-3/2 particles we then use these interactions to 
on-shell construct the four-point massive spin-3/2 amplitude. To do this, we rely on the novel, all-line-transverse (ALT) shift introduced in Refs.~\cite{Ema:2024rss,Ema:2024vww} which was used for spin-1/2 amplitudes in QED and massive gauge boson amplitudes in electroweak theory. When applied to massive spin-3/2 fields, we find that the four-point amplitude for transverse states is on-shell constructible from graviton exchange, 
provided the Ward identity is imposed. Interestingly, the four-point gravitino contact interaction, that appears in the $N=1$ supergravity Lagrangian, arises from the graviton exchange under the ALT shift. 
This four-point amplitude is then used to check known constraints on the coupling coefficients that arise from perturbative unitarity. Without using a Lagrangian or Feynman diagrams, we recover the well-known result that by adding a scalar and a pseudoscalar interaction the high-energy behavior scales as $E^2$ with a Planck scale unitarity bound, which corresponds to the (on-shell) Polonyi model. 

As a final check of the on-shell approach, we consider the high-energy limit of the massive scattering amplitude (computed using the ALT shift) corresponding to the process $ \frac{3}{2}, -\frac{3}{2} \rightarrow \frac{1}{2},-\frac{1}{2}$. By the equivalence theorem, the high-energy limit is equivalent to the {\it massless} scattering process involving a massless spin-3/2 and a massless spin-1/2 fermion (associated with the longitudinal component of the massive spin-3/2 particle). Since all external states are now massless, we use an all-line shift, which has good large-$z$ behavior, to compute the scattering amplitude via the on-shell recursion relation. As expected the two results match where, again, no Lagrangians or Feynman diagrams were used in the calculation.

The outline of this paper is as follows. In section~\ref{sec:spinorhel}, we review the massive spinor-helicity formalism as applied to spin-3/2 particles. This is then used to construct the various on-shell three-point amplitudes involving spin-3/2 particles in section~\ref{sec:trilinear}. A four-point massive spin-3/2 amplitude is calculated in section~\ref{secfourpoint} where we define the all-line transverse shift for spin-3/2 particles, compute a massive gravitino scattering amplitude and discuss the high-energy limit. Our concluding remarks are given in section~\ref{sec:concl}.
Some background material and further details are given in the appendices. This includes conventions and useful relations in appendix~\ref{app:conv}, explicit kinematic formulae in appendix~\ref{app:kin}, details of the constraints from the current conservation calculation in appendix~\ref{app:gaugeinv} and a review of the all-line-transverse shift in appendix~\ref{app:recursion}.

\newpage
 \section{Massive spinor-helicity formalism} 
 \label{sec:spinorhel}
 
In this section, we present the essential elements of the massive spinor-helicity formalism. It is based on the massless spinor helicity formalism where the momentum bispinor $p_{\alpha\dotalpha}=p_\mu \sigma^\mu_{\alpha\dotalpha}=\lambda_\alpha\tilde{\lambda}_{{\dot{\alpha}}}\equiv\left|p\right>\left[p\right|$, with four-momentum $p_\mu$ and Pauli matrices $\sigma^\mu$, is decomposed into a pair of chiral $(\lambda_\alpha)$ and anti-chiral $(\tilde{\lambda}_{{\dot{\alpha}}})$ spinors. In our notation, the spinors with dotted and undotted indices transform respectively under the $(0,1/2)$, and $(1/2,0)$ representations of the $SL(2,\mathbb{C})$ group. More details and conventions can be found in appendix \ref{app:conv}.

The massless spinor formalism can be generalized to the case of massive particles. Following Ref.~\cite{Arkani-Hamed:2017jhn}, the massive momentum bispinor, satisfying $p^2=m^2$, is now a rank-two  matrix which can then be decomposed into the spinor variables
\begin{equation}
 {p}_{\alpha\dotalpha}= {\lambda}^I_\alpha \tilde{  {\lambda}}_{I \dotalpha  }\equiv\left| \boldsymbol{p}^I\right>\left[ \boldsymbol{p}_I\right|\,,
 \label{momentummassive}
\end{equation}
where $I=1,2$ are the $SU(2)$ little-group indices that are raised and lowered by $\varepsilon^{IJ}$, $\varepsilon_{IJ}$ with $\varepsilon_{12}=-1$. The massive  spinor variables in the \textit{bra-ket} notation are denoted by bold symbols and our conventions are summarized in appendix \ref{app:conv}.

The little-group transformations that leave the massive momentum invariant are:
\begin{equation} {\lambda} ^I\rightarrow W^I_J \lambda^J,\quad  \tilde{\lambda} _I\rightarrow (W^{-1})_I^J \tilde{\lambda}_J\,,
\end{equation}
where $W$ is an $SU(2)$ matrix. Since we will be interested in studying the high-energy limit and defining a massive momentum shift, 
it is convenient to decompose the massive spinors into a basis with massless spinors, namely
\begin{equation}   
\begin{aligned}        
&\lambda^I_\alpha=\lambda_\alpha\zeta^{-I}+\eta_\alpha\zeta^{+I }\,, \\&\Tilde{\lambda}^I_{\dotalpha}=\Tilde{\lambda}_{\dot\alpha}\zeta^{+I}+\Tilde{\eta}_{\dotalpha}\zeta^{-I }\,,   
\end{aligned}
\label{decomp-lam}
\end{equation}
which satisfies 
\begin{equation}   \zeta^{-I}\zeta^+_I=1,\quad  \lambda^\alpha\eta_\alpha = \Tilde{\lambda}_{\dotalpha}\Tilde{\eta}^{\dotalpha}=m\,,
\end{equation}
and $\lambda, \Tilde{\eta}$ have $-1/2$ helicity, whereas $\Tilde{\lambda}, \eta$ have $+1/2$ helicity. Therefore, contrary to the massless case, the appearance of a spinor  
in a square or angle bracket is no longer uniquely determined by the helicity, since both angle and square brackets are possible. We write $|\bp)$ when the helicity is unspecified. The massive momentum \eqref{momentummassive} is re-expressed as
\begin{equation}
    p_{\alpha\dotalpha}=\lambda_\al\Tilde{\lambda}_{\dotalpha}-\eta_\al\Tilde{\eta}_{\dotalpha}\,.
\end{equation} 
The explicit kinematics of $\lambda$,  $\Tilde{\lambda}$, $\eta$, $\Tilde{\eta}$ are given in appendix \ref{app:conv}, from which we can see that at high energies ($E\gg m$), they scale as $\lambda,\Tilde{\lambda}\sim \sqrt{E}$, $\eta,\Tilde{\eta}\sim \frac{m}{\sqrt{E}}$.

A particle of spin $s$ can be obtained from $2s$ chiral ($\lambda$) or anti-chiral ($\tilde{\lambda}$) spinors.  The spin-1/2 and spin-1 polarization vectors are presented in appendix\,\ref{app:conv}, and we review here the construction of a massive spin-3/2 particle which is essential for obtaining our results. The spin-3/2 particle can be constructed from the tensor product of Dirac and vector representations, so its polarization vector is 
\begin{equation}
    \bar{v}_\mu^{IJK}(\bp)=\left(\begin{array}{cc}
        \left<\bp\right| \bar{\varepsilon}_\mu& \left[\bp\right| \bar{\varepsilon}_\mu 
    \end{array}\right),\quad   u_\mu^{IJK}(\bp)=\left(\begin{array}{c}
         \left|\bp\right> \varepsilon_\mu \\
\left|\bp\right]\varepsilon_\mu
    \end{array}\right)\,.
    \label{eq:polvec32}
\end{equation}
where $\varepsilon_\mu$ is the massive  spin-1 polarization and we have dropped the little-group 
indices $I,J,K$ on the right-hand side.
It is by construction divergence-free because $p^\mu\varepsilon_\mu=0$ and  gamma-traceless because $\left<\bp\bp\right>^{(IJ)}=\left[\bp\bp\right]^{(IJ)}=0$ (see appendix \ref{app:conv}). The massive spin-3/2 particle is decomposed into different helicities weighted with Clebsch-Gordan coefficients
\begin{equation}\begin{aligned}
    &\psi_\mu^{++}=\varepsilon_\mu^+ v^+ ,\quad \psi_\mu^{--}=\varepsilon_\mu^- v^-,\\&\psi_\mu^+=\sqrt{\frac{2}{3}}\varepsilon_\mu^0 v^++\sqrt{\frac{1}{3}}\varepsilon_\mu^+ v^-
,\quad \psi_\mu^-=\sqrt{\frac{2}{3}}\varepsilon_\mu^0 v^-+\sqrt{\frac{1}{3}}\varepsilon_\mu^-v^+\,,
\end{aligned}
\label{gravitino-helicities}\end{equation}
where double indices $\pm\pm$ denote $\pm 3/2$ helicities  and single indices $\pm$ are  $\pm1/2$ helicities. 

In this work, we will be interested in Majorana fermions. A Majorana fermion can be obtained by imposing a reality condition on a Dirac fermion. In our convention, all particles are incoming with momentum conservation $\sum_i p_i=0$. Hence, the spin-3/2 $\psi_\mu$ and its conjugate $\psib_\mu$ will be represented by $u_\mu(p)$ and $\bar{v}_\mu(p)$, respectively.

Having set up the particle representations, we now turn to amplitudes. The amplitude containing a spin-$s$ particle can then be written as \cite{Arkani-Hamed:2017jhn}
\begin{equation}
    M^{I_1\cdots I_{2s}}=\lambda_{\alpha_1}^{I_1}\cdots \lambda_{\alpha_{2s}}^{I_{2s}}M^{\alpha_1\cdots \alpha_{2s}}\,,
\end{equation}
where $\{\alpha_1,\cdots ,\alpha_{2s}\}$ are totally symmetric and $\{I_1,\cdots, I_{2s}\}$ are little-group indices. The above notation allows for a systematic classification of the little-group covariant amplitude structures, for given masses and spins.  Each massive external particle of spin $s_i$ will appear in the amplitude with $2s_i$ \textit{symmetrized} (with appropriate normalization coefficients) little-group indices $\{I,J,K,...\}=1,2$. In section~\ref{sec:trilinear}, we  review more details about all possible three-point amplitudes and apply them to spin $3/2$, following the prescription of \cite{Arkani-Hamed:2017jhn}. As long as there is no ambiguity, the little-group indices will be dropped for simplicity.

 \section{On-shell three-point couplings}
 \label{sec:trilinear}

The spinor-helicity formalism developed in \cite{Arkani-Hamed:2017jhn} provides a systematic classification of all possible three-point amplitudes consistent with little-group and Lorentz group transformations. From the amplitudes expressed in spinor variables, we can infer the corresponding three-point couplings. This is similar to how the on-shell formalism has been applied to construct  effective operators in the Standard Model   \cite{Christensen:2018zcq,Durieux:2019eor,Christensen:2019mch,Goldberg:2024eot}. 

We will consider effective three-point couplings involving a massive spin-3/2 particle coupled to gravity, along with scalar particles, a spin-1/2 fermion and  a massless gauge boson. Unlike the electroweak case, there is now a fundamental mass scale in the theory, the Planck mass $M_P$, allowing for   non-renormalizable interactions with a coupling $\kappa\equiv 2/M_P$.  The latter appears, for example, when the metric is expanded around the flat Minkowski background $g_{\mu\nu}=\eta_{\mu\nu}+\kappa h_{\mu\nu}+\cdots$, where $h_{\mu\nu}$ represents the graviton.  The particle content and the masses are summarized below:

\begin{center}
\begin{tabular}{|c|c|c|}
\hline
& spin & mass\\
\hline\hline
scalar, $S$ & 0 &$m_S$\\
pseudoscalar, $P$ & 0 &$m_P$\\
fermion, $\chi$ & $\frac{1}{2}$ & $m_\chi$\\
gauge boson, $A_\mu$ & 1 & $0$\\
\hline\hline
spin 3/2, $\psi_\mu$ & $\frac{3}{2}$ & $m_{3/2}$\\
graviton, $h_{\mu\nu}$& 2& $0$\\
\hline
\end{tabular}
\end{center}
The gauge group associated with $A_\mu$ can be Abelian or non-Abelian.  
To simplify our discussion we will not specify the gauge group and just present the three-point amplitudes with the color factors stripped off.

We will enumerate all possible three-point amplitudes that contain either one or two massive spin-3/2 particles, interacting with the above particle content. The amplitudes involving two spin 3/2 can interact with one spin $s=0,1,2$, while those involving one spin 3/2 can necessarily only interact with one spin 1/2 together with one spin $s=0,1,2$. 
Also note that the spin 3/2 is assumed to be a \textit{Majorana} fermion, which imposes antisymmetry of the amplitude under the exchange of two identical spin-3/2 particles.

\subsection{$(s,3/2, 3/2)$}\label{s3232}

The basis for the on-shell three-point amplitudes involving two massive spin-3/2 particles, and another massive particle is given in Table 2 of \cite{Durieux:2020gip}. We extend this basis by considering interactions of spin-3/2 particles with massless states of spin $s\leq 2$ 
and reduce the basis by the Majorana condition as well as the requirement that the amplitude arise from interactions of dimension $\leq6$.  We then match each spinor form of the amplitude to examples of interactions in the Lagrangian.

The spin $s$ particle is labelled as $\bone$, while the two spin-3/2 particles are labelled as $\btwo,\bthree$.   Since a pair of fermions can only couple to a boson, $s$ is an integer. Moreover, Fermi-Dirac statistics require that the amplitude be antisymmetric under the   exchange $\btwo\leftrightarrow \bthree$. 
Also, in the spinor-helicity language, parity  flips angle brackets and square brackets. In particular, parity sends $\left<\btwo p_1 \bthree\right]$ to $\left<\bthree p_1 \btwo\right]$ and $\left<\btwo \varepsilon^+_1 \bthree\right]$ to $\left<\bthree\varepsilon^-_1 \btwo\right]$ which will be used to distinguish parity-even and parity-odd amplitudes.

\subsubsection{Massive spin $s$}

When all external legs are massive, the most general little-group covariant three-point amplitude can be constructed by combining $2s$ $\left|\bone\right)$, three $\left|\btwo \right)$ and three $\left|\bthree \right)$ brackets, 
as well as $\varepsilon_{\alpha \dotalpha}$ matrices  to raise and lower spinor indices \cite{Arkani-Hamed:2017jhn,Durieux:2019eor}. Furthermore, one may insert momenta to obtain other independent structures. In our setup, the only massive boson in the particle content is the  (pseudo-)scalar, therefore we restrict ourselves to $s=0$, though the result can be easily generalized to  higher spins (see, for example,~\cite{Durieux:2020gip}). 

There are four spinor structures and all of them are antisymmetric under $\btwo\leftrightarrow \bthree$:
\begin{equation}
    \left[\btwo\bthree\right]^3,\quad \left<\btwo \bthree\right>^3,\quad\left<\btwo\bthree\right>^2\left[\btwo\bthree\right],\quad\left<\btwo\bthree\right>\left[\btwo\bthree\right]^2\,.
    \label{basis-phipsipsi}
\end{equation}
 Note that any momentum insertion in \eqref{basis-phipsipsi} does not yield an independent structure, upon using the momentum conservation $p_1+p_2+p_3=0$ and Dirac equations. As for the high-energy behavior, the first and second terms in \eqref{basis-phipsipsi} scale as $\mathcal{O}(E^3)$, in the helicity amplitudes $\mathcal{M}^{+\frac{3}{2},+\frac{3}{2},0}$, and $\mathcal{M}^{-\frac{3}{2},-\frac{3}{2},0}$ respectively. On the other hand, the dominant energy behavior of the last two structures is
\begin{equation}   \left<\btwo\bthree\right>^2\left[\btwo\bthree\right]\rightarrow\left\{ \begin{aligned}
    &\frac{1}{\sqrt{3}}\left<23\right>^2 \left[\Tilde{\eta}_2 3\right]=-\frac{\left<23\right>^2\left<12\right>}{\sqrt{3}\left<31\right>}m_{3/2}\sim\mathcal{O}(E^2)\quad\text{for }\left(0,-\frac{3}{2},-\frac{1}{2}\right)\,,
    \\&\frac{1}{\sqrt{3}}\left<23\right>^2 \left[2\Tilde{\eta}_3\right]=\frac{\left<23\right>^2\left<31\right>}{\sqrt{3}\left<12\right>}m_{3/2}\sim\mathcal{O}(E^2)\quad\text{for }\left(0,-\frac{1}{2},-\frac{3}{2}\right)\,,
    \end{aligned}\right.
\label{masslesspsipsis1}\end{equation}where the factor $1/\sqrt{3}$ accounts for the normalization. 
Since $\left<\btwo\bthree\right>\left[\btwo\bthree\right]^2$ is related to $\left<\btwo\bthree\right>^2\left[\btwo\bthree\right]$ by parity, we immediately obtain 
\begin{equation}
    \left<\btwo\bthree\right>\left[\btwo\bthree\right]^2=\left\{ \begin{aligned}
    &\frac{1}{\sqrt{3}}\left[23\right]^2 \left<{\eta}_2 3\right>=- \frac{\left[23\right]^2\left[12\right]}{\sqrt{3}\left[31\right]}m_{3/2}\sim\mathcal{O}(E^2)\quad\text{for }\left(0,+\frac{3}{2},+\frac{1}{2}\right)\,,
    \\&\frac{1}{\sqrt{3}}\left[23\right]^2 \left<2{\eta}_3\right>=\frac{\left[23\right]^2\left[31\right]}{\sqrt{3}\left[12\right]}m_{3/2}\sim\mathcal{O}(E^2)\quad\text{for }\left(0,+\frac{1}{2},+\frac{3}{2}\right)\,.
    \end{aligned}\right.
    \label{masslesspsipsis2}
\end{equation}
It is straightforward to see that the massless limit of the massive amplitudes correctly reproduces Eqs.~\eqref{masslesscplus}-\eqref{masslesscminus} up to global coupling coefficients. Indeed, this matching can be interpreted on-shell as the super-Higgs mechanism \cite{Arkani-Hamed:2017jhn} --  starting with a gravity multiplet $(h_{\mu\nu},\psi_\mu)$ and a  chiral  multiplet $(\phi,\chi)$ ($\chi$ is  a spin-1/2 fermion), the universal $\psi_\mu \chi \phi$ coupling is uplifted to a massive spin-3/2 coupling to a scalar, thereby recovering a massive spin-3/2 multiplet.

The next order in the expansion at $\mathcal{O}(E)$ corresponds to the $\left(0,\pm 1/2, \pm 1/2 \right)$ helicity amplitudes
\begin{equation}
    \begin{aligned}
  &\left<\btwo\bthree\right>\left[\btwo\bthree\right]^2
\rightarrow-\frac{1}{3}m_{3/2}^2 [23]\quad\text{for }\left(0,+\frac{1}{2},+\frac{1}{2}\right)\,,
\\& \left<\btwo\bthree\right>^2\left[\btwo\bthree\right] 
\rightarrow-\frac{1}{3}m_{3/2}^2 \left<23\right>\quad\text{for }\left(0,-\frac{1}{2},-\frac{1}{2}\right)\,.
\end{aligned}
\end{equation} 
They can be seen as  Yukawa couplings of the longitudinal modes. 

To derive examples of Lagrangian interactions that correspond to the amplitudes in \eqref{basis-phipsipsi}, recall that the spin-3/2 polarization vector is a combination of $\left|\boldsymbol{i}\right>\left|\boldsymbol{i}\right>\left|\boldsymbol{i}\right]$ and $\left|\boldsymbol{i}\right]\left|\boldsymbol{i}\right]\left|\boldsymbol{i}\right>$ ($\boldsymbol{i}=\btwo,\bthree$). Hence, for an amplitude containing only square or angle brackets, there is inevitably at least one momentum (derivative) insertion to ``exchange'' angle and square through the Dirac equation.  In particular, $ \left[\btwo\bthree\right]^3, \left<\btwo \bthree\right>^3$ may arise from the dimension-six coupling $\phi \partial_\mu\psib_\nu\partial_\rho\psi_\sigma $ up to contractions with $\gamma$ matrices or the metric ($\phi$ schematically represents the scalar $S$ or the pseudoscalar $P$).   
On the other hand, $\left<\btwo\bthree\right>^2\left[\btwo\bthree\right]$, $ \left<\btwo\bthree\right>\left[\btwo\bthree\right]^2$ can be obtained   by an operator  of the form $\phi \psib^\mu\psi_\mu $, without any derivatives. 

With different linear combinations one can build parity-even or parity-odd amplitudes. Leaving for now the coupling constants unspecified, let us summarize the typical dimension $\leq6$ operators that can be related to these linear combinations:
\begin{equation}
\begin{aligned}\text{Parity even: }&\left< \btwo\bthree\right>^2\left[\btwo\bthree\right]+\left<\btwo\bthree\right>\left[\btwo\bthree\right]^2\quad\longrightarrow\quad \psib^\mu\psi_\mu S\\ &\left(\left<\btwo\bthree\right>+\left[\btwo\bthree\right]\right)\left(\left<\btwo\bthree\right>^2+\left[\btwo\bthree\right]^2\right)\quad\longrightarrow\quad  \partial^{[\mu}\psib^{\nu]}\partial_{[\mu}\psi_{\nu]}S \\\text{Parity odd: }& \left< \btwo\bthree\right>^2\left[\btwo\bthree\right]-\left<\btwo\bthree\right>\left[\btwo\bthree\right]^2\quad \longrightarrow\quad \varepsilon^{\mu\nu\alpha\beta } \partial_\mu P \psib_\nu \gamma_\alpha\psi_\beta\\ &\left(\left<\btwo\bthree\right>+\left[\btwo\bthree\right]\right)\left(\left<\btwo\bthree\right>^2-\left[\btwo\bthree\right]^2\right)\quad \longrightarrow\quad \varepsilon^{\mu\nu\alpha\beta } P  \partial_\mu\psib_\nu \partial_\alpha\psi_\beta
\end{aligned}
\label{phipsipsiinteraction}
\end{equation}
Note that there are other interactions that vanish on-shell, for example $\partial^\mu\psib_\mu\partial^\nu\psi_\nu \phi$, which would be relevant in the Lagrangian in order to derive the equations of motion.
These off-shell interactions in the Lagrangian increase the complexity of studying spin-3/2 interactions which are absent in the on-shell formulation. 
We also omit interactions that are on-shell equivalent  to those listed in \eqref{phipsipsiinteraction}. For instance, $\psib_\mu\gamma^{\mu\nu}\psi_\nu S$ amounts to $\psib^\mu\psi_\mu S$ by using the trace constraint (see \eqref{eq:constraints}).
Moreover, throughout this work, we assign $S$ to parity even interactions and $P$ to parity odd ones, which is just a convention. Typically, an interaction of the form $\psib^\mu \gamma^5\psi_\mu S$ is parity odd but is not considered here. Finally, note that the first- and third-line interactions of \eqref{phipsipsiinteraction} feature in  $N=1$ supergravity  which arise from the gravity supermultiplet coupled to a chiral matter multiplet.

\subsubsection{Massless spin $s$}
The general little-group covariant three-point amplitude with one massless state of helicity $h$,  and two massive particles with equal mass is presented in \cite{Arkani-Hamed:2017jhn}. 
By factoring out the massive spinors, the amplitude in the case of spin 3/2 can be written as 
\begin{equation}
    M^{\{I_1I_2I_3J_1J_2J_3\},h}=\lambda_2^{I_1\alpha_1}\lambda_2^{I_2\alpha_2}\lambda_2^{I_3\alpha_3}\lambda_3^{J_1\beta_1}\lambda_3^{J_2\beta_2}\lambda_3^{J_3\beta_3}M^h_{\{\alpha_1\alpha_2\alpha_3\},\{\beta_1\beta_2\beta_3\}}\,.
    \label{nima-1}
\end{equation}
Recall that the labels 2, 3 represent the spin-3/2 particles, while label 1 is the massless particle.  For equal mass particles, one introduces an ``$x$-factor'' which encodes the proportionality between $\lambda_1$ and $\Tilde{\lambda}_1$:
\begin{equation}
    x\lambda_{1\alpha}=\frac{p_{2\al\dotalpha}}{m}\Tilde{\lambda}_1^{\dotalpha},\qquad \frac{\tilde{\lambda}_{1}^{\dotalpha}}{x}=\frac{p_2^{\dotalpha\alpha}}{m}
    \lambda_{1\alpha}\,,
    \label{xfactordef}
\end{equation} 
where $m=m_{3/2}$ is the spin-3/2 mass in our case. 
The three-point amplitude can then be written as a power series associated with coupling constants $g_i$
\begin{equation}
    M^h_{\{\alpha_1\alpha
    _2\alpha_3\},\{\beta_1\beta_2\beta_3\}}=\sum_{i=0}^3g_ix^h\left[\lambda_1^i\left(\frac{p_2\Tilde{\lambda}_1}{m}\right)^i\varepsilon^{3-i}\right]_{\{\alpha_1\alpha
    _2\alpha_3\},\{\beta_1\beta_2\beta_3\}}\,.
    \label{nima-3}
\end{equation}
The first term in this series is  called the \textit{minimal coupling} in the sense that, if the massless particle is a photon, the  gyromagnetic ratio is $g=2$ \cite{Chung:2018kqs} \footnote{Note however that in the charged higher-spin literature, minimal coupling usually designates replacing partial derivatives $\partial_\mu$ by covariant derivatives $\mathcal{D}_\mu=\partial_\mu -i e A_\mu$, which in general does not lead to $g=2$.}, whereas the higher-order terms correspond to higher multipoles.  Putting back the massive spinors, the minimal coupling is given  by \cite{Arkani-Hamed:2017jhn,Chung:2018kqs}
\begin{equation}
    (mx)^h\left(\frac{\left<\btwo\bthree\right>}{m}\right)^{2s}\,.
    \label{mincoup-x}
\end{equation} 
In the following, we show the results for massless spin 1 and spin 2, where some useful relations involving the $x$-factor are listed in appendix \ref{app:conv}, Eq.~\eqref{xrelation}-\eqref{polarelation}.

\begin{itemize}
\item $s=1$
 
Applying Eqs.~\eqref{nima-1}-\eqref{nima-3}, the amplitude for a positive helicity $(+1)$ gauge boson is
\begin{equation}\begin{aligned}
    M^+=&\,g_0^{(+1)} x\left<\btwo\bthree\right>^3+g_1^{(+1)}\frac{x}{m_{3/2}}\left<\btwo\bthree\right>^2\left<\bthree|p_2|1\right]\left<\btwo 1\right>+g_2^{(+1)} \frac{x}{m_{3/2}^2} \left<\btwo\bthree\right>\left<\btwo1\right>^2\left<\bthree|p_2|1\right]^2\\
    &+g_3^{(+1)}\frac{x}{m_{3/2}^3}\left<\btwo1\right>^3\left<\bthree|p_2|1\right]^3\,,
    \\  =&\left(\left<\btwo\varepsilon^+_1\bthree\right]+\left<\bthree\varepsilon^+_1\btwo\right]\right)\\&\times\left[g_0^{(+1)} \left<\btwo\bthree\right>^2  -g_1^{(+1)}m_{3/2}\left<\btwo\bthree\right>\left[\btwo\bthree\right]+g_1^{(+1)}m_{3/2}\left<\btwo\bthree\right>^2+g_2^{(+1)} m_{3/2}^2 \left( \left[\btwo\bthree\right]- \left<\btwo\bthree\right>\right)^2\right]\\&-g_3^{(+1)} m_{3/2}^2 \left[1\btwo \right] \left[1\bthree \right]\left( \left[\btwo\bthree\right]- \left<\btwo\bthree\right>\right)^2\,, 
\end{aligned}\label{gravphoton}
    \end{equation}
where in the second line we have used \eqref{xrelation}. A remarkable feature, also noticed in \cite{Durieux:2019eor} for the case of spin-1/2 fermion coupling to a photon, is that the spinor-helicity formalism constrains the coupling to be purely vector-like: the factor $\left(\left<\btwo\varepsilon^+_1\bthree\right]+\left<\bthree\varepsilon^+_1\btwo\right]\right)$ arises from $\psib^\mu \slashed{A} \psi^\nu$ whereas the opposite relative sign $ \left(\left<\btwo\varepsilon^+_1\bthree\right]-\left<\bthree\varepsilon^+_1\btwo\right]\right) $  is associated with the axial-vector coupling $\psib^\mu \slashed{A} \gamma^5\psi^\nu$. 

However, because we have assumed the spin-3/2 particle to be  Majorana, the Fermi-Dirac statistics imposes antisymmetry under $\btwo\leftrightarrow\bthree$ which is satisfied by neither of the above terms. Therefore, we conclude that the Majorana spin-3/2 particle cannot be charged and the coupling vanishes. More generally, it would be interesting to consider charged, massive spin-3/2 particles which are Dirac fermions. In this case, Eq.~\eqref{gravphoton} would be useful for finding the possible Lagrangian interactions.

\item $s=2$

The graviton polarization vector is simply the product of two massless spin-1 polarization vectors
\begin{equation}
\varepsilon_{\mu\nu}^+=\varepsilon_\mu^+\varepsilon^+_\nu,\quad \varepsilon^-_{\mu\nu}=\varepsilon^-_\mu\varepsilon^-_\nu \,.   
\end{equation}
 The three-point amplitude for the helicity $+2$ graviton is proportional to  Eq.~\eqref{gravphoton} multiplied by an $x$-factor. While Eq.~\eqref{gravphoton} is symmetric under $\btwo\leftrightarrow\bthree$, the extra $x$-factor brings a minus sign under this exchange because $0=\left<\zeta|p_1|1\right]+\left<\zeta|p_2|1\right]+\left<\zeta|p_3|1\right]=\left<\zeta|p_2|1\right]+\left<\zeta|p_3|1\right]$. The resulting amplitude therefore satisfies Fermi-Dirac statistics. 
 
We can write for the helicity $+2$ graviton
\begin{equation}
\begin{aligned}
 M^{+2}=&\,g_0^{(+2)} x^2\left<\btwo\bthree\right>^3+g_1^{(+2)}\frac{x^2}{m_{3/2}}\left<\btwo\bthree\right>^2\left<\bthree|p_2|1\right]\left<\btwo 1\right>+g_2^{(+2)} \frac{x^2}{m_{3/2}^2} \left<\btwo\bthree\right>\left<\btwo1\right>^2\left<\bthree|p_2|1\right]^2\\&+g_3^{(+2)}\frac{x^2}{m_{3/2}^3}\left<\btwo1\right>^3\left<\bthree|p_2|1\right]^3\,, \\       =&\left(\left<\btwo\varepsilon^+_1\bthree\right]+\left<\bthree\varepsilon^+_1\btwo\right]\right)^2\left (g_0^{(+2)} \left<\btwo\bthree\right>+g_1^{(+2)} \left<1\btwo \right>[1\bthree]\right)+g_2^{(+2)}[1\btwo]^2[1\bthree]^2 \left<\btwo \bthree\right>\\&+g_3^{(+2)} \left<1\btwo \right>[1\btwo]^2[1\bthree]^3\,, \\       =&\left(\left<\btwo\varepsilon^+_1\bthree\right]+\left<\bthree\varepsilon^+_1\btwo\right]\right)^2\left({\hat g}_0^{(+2)} \left<\btwo\bthree\right> +{\hat g}_1^{(+2)} [\btwo\bthree]\right) 
 + {\hat g}_2^{(+2)} \left<\btwo\bthree\right>[1\btwo]^2 [1\bthree]^2\\&+{\hat g}_3^{(+2)} [\btwo\bthree ][1\btwo]^2 [1\bthree]^2\,.
\end{aligned}
\label{gravitonplus}
\end{equation}
In the last step, we have redefined the coupling coefficients to absorb identical terms, with
\begin{equation}
    {\hat g}_{0,2}^{(+2)}=g_{0,2}^{(+2)} +m_{3/2} g_{1,3}^{(+2)},\quad {\hat g}_{1,3}^{(+2)} =-m_{3/2} g_{1,3}^{(+2)}\,. 
\end{equation}
Notice that the ${\hat g}_2^{(+2)}$, ${\hat g}_3^{(+2)}$ terms correspond to dimension 6 and dimension 8 operators, respectively. Similarly, for the helicity $-2$ graviton
\begin{equation}
    \begin{aligned}
        M^{-2} =&\left(\left<\btwo\varepsilon^-_1\bthree\right]+\left<\bthree\varepsilon^-_1\btwo\right]\right)^2\left({\hat g}_0^{(-1)} [\btwo\bthree]+{\hat g}_1^{(-1)} \left<\btwo\bthree\right>\right) + {\hat g}_2^{(-1)} \left[\btwo\bthree\right]\left<1\btwo\right>^2 \left<1\bthree\right>^2\\
        &+ {\hat g}_3^{(-1)} \left<\btwo\bthree \right>\left<1\btwo\right>^2 \left<1\bthree\right>^2\,.
    \end{aligned} 
    \label{psipsihminus}
    \end{equation} 
The on-shell couplings \eqref{gravitonplus}, \eqref{psipsihminus} have several interesting implications. For example, the  dimension-four term $h_{\mu\nu}\psib^\mu\psi^\nu$ yields the amplitude
$\left<\btwo\varepsilon_1^\pm \bthree\right]\left<\bthree\varepsilon_1^\pm \btwo\right]\left(\left<\btwo\bthree\right>+\left[\btwo\bthree\right]\right)$ which is not an independent term in \eqref{gravitonplus}. This means that $h_{\mu\nu}\psib^\mu\psi^\nu$ cannot appear alone, and other (derivative) couplings, with non-vanishing on-shell amplitudes, are necessary to make up    the squared prefactor $\left(\left<\btwo\varepsilon^\pm_1\bthree\right]+\left<\bthree\varepsilon^\pm_1\btwo\right]\right)^2$. 

Furthermore, we can translate the three-point amplitudes \eqref{gravitonplus}, \eqref{psipsihminus} into possible Lagrangian interactions to compare with the usual off-shell formulation. A well-known example is the  gravitational minimal coupling
\begin{equation}
    \frac{\kappa }{2}h_{\mu\nu}T_{3/2}^{\mu\nu}\,,
    \label{mincoup32}
\end{equation}
where $T_{3/2}^{\mu\nu}$ is the energy momentum tensor obtained by varying   the Rarita-Schwinger action with regard to the linear perturbation of the metric, $g_{\mu\nu}\simeq\eta_{\mu\nu}+\kappa h_{\mu\nu}$. Off-shell, this coupling reads,\begin{equation}
\begin{aligned}
   \kappa h_{\mu\nu }T^{\mu\nu}_{3/2}=&\frac{\kappa}{4}\left( h \varepsilon^{\rho\beta \mu\nu}\psib_\rho \gamma_5 \gamma_\beta \overset{\leftrightarrow}{\partial_\mu} \psi_\nu-\frac{1}{2} h_{\rho\tau }\varepsilon^{\mu\nu\rho\sigma}\partial_\lambda \left(\psib_\mu\gamma_5 \gamma_\sigma\gamma^{\tau\lambda}\psi_\nu \right)\right.\\
   &\left.- h_{\rho\tau }\varepsilon^{\mu\nu\rho\lambda}\psib_\mu \gamma_5 \gamma^\tau \partial_\lambda\psi_\nu + m_{3/2}  h  \psib_\mu \gamma^{\mu\nu }\psi_\nu + m_{3/2} h ^{\mu\alpha }\psib_\mu\gamma_{\alpha\nu } \psi^\nu\right.\\
   & \left.+ m_{3/2} h ^{\nu\alpha }\psib^\mu\gamma_{\mu\alpha } \psi_\nu\right)\,,
\end{aligned}
\label{gravitinotmunu}
\end{equation}
where $A \overset{\leftrightarrow}{\partial_\mu} B\equiv A(\partial_\mu B)-(\partial_\mu  A)B$, and $h\equiv g^{\mu\nu}h_{\mu\nu}$.
When the gravitino  and the graviton are on-shell, \eqref{gravitinotmunu} becomes 
\begin{equation} 
\kappa h_{\mu\nu} \left. T_{3/2}^{\mu\nu}\right|_\text{on-shell}= i\frac{\kappa}{2}h_{\mu\nu}\left(\psib^\rho \gamma^\mu \partial_\rho \psi^\nu -\partial_\rho \psib^\mu \gamma^\nu \psi^\rho -\frac{1}{2}\psib_\rho \gamma^\mu \overset{\leftrightarrow}{\partial^\nu} \psi^\rho\right)\,. 
\label{mincouponshell}
\end{equation}  
Using the relations \eqref{polarelation}, one can show that the minimal coupling \eqref{mincouponshell} corresponds to
\begin{equation}
    \kappa h_{\mu\nu}T_{3/2}^{\mu\nu}=\left\{ \begin{aligned}
    &-\frac{\kappa}{4m_{3/2}}\left<\btwo\bthree\right>\left(\left<\btwo\varepsilon^+_1\bthree\right]+\left<\bthree\varepsilon^+_1\btwo\right]\right)^2\,,
    \\&-\frac{\kappa}{4m_{3/2}}\left[\btwo\bthree\right]\left(\left<\btwo\varepsilon^-_1\bthree\right]+\left<\bthree\varepsilon^-_1\btwo\right]\right)^2\,,
    \end{aligned}\right.
    \label{mincoupinspinor}
  \end{equation}
implying ${\hat g}_0^{(+2)}=g_0^{(+2)} \neq0$, ${\hat g}_1^{(+2)} ={\hat g}_2^{(+2)} ={\hat g}_3^{(+2)} =0$ for the positive helicity graviton, and ${\hat g}_0^{(-1)}  \neq0$, ${\hat g}_1^{(-1)} ={\hat g}_2^{(-1)} ={\hat g}_3^{(-1)} =0$ for the negative helicity graviton, which  precisely coincides with the ``minimal coupling'' defined in \eqref{mincoup-x}. In other words, the gravitational minimal coupling can be identified with the first term in the series \eqref{nima-3} which does not have a 
gravitational dipole coupling. 
In the general case with any spin, the matching of the minimal couplings was proven in Ref.~\cite{Chung:2018kqs}. 
  
In fact, notice that the first two contributions in \eqref{mincouponshell} and the third term separately have a dipole component. They correspond to the following three-point amplitudes
\begin{equation}
    \begin{aligned}
    &{\hat g}_1^{(+2)}\neq 0,\quad {\hat g}_0^{(+2)} ={\hat g}_2^{(+2)} ={\hat g}_3^{(+2)}=0\quad\longrightarrow\quad i h_{\mu\nu}\psib_\rho
    \gamma^\mu\partial^\nu\psi^\rho +\text{h.c.}\\
    &{\hat g}_0^{(+2)} =-{\hat g}_1^{(+2)}\neq0 , \quad{\hat g}_2^{(+2)} ={\hat g}_3^{(+2)} =0\quad\longrightarrow\quad ih_{\mu\nu}\psib^\rho \gamma^\mu\partial_\rho\psi^\nu +\text{h.c.}
    \label{eq:3ptdipoleint}
\end{aligned}
\end{equation}  
Therefore, as can be seen in \eqref{mincouponshell}, the relative coefficients of the two interactions in \eqref{eq:3ptdipoleint} are uniquely fixed,  
which end up cancelling the gravitational dipole coupling.\,\footnote{The absence of a gravitational dipole coupling for a consistent theory of any spin and any mass was argued, for example, in   \cite{Chung:2018kqs,Falkowski:2020aso}.   This requirement actually imposes $g_1^{(+2)}=0$, hence ${\hat g}_0^{(+2)} =g_0^{(+2)}$, ${\hat g}_1^{(+2)}=0$.  Beyond  the dipole,  some constraints have also been  derived for  the gravitational quadrupole moment   \cite{Giannakis:1998wi,Porrati:1993in,Cucchieri:1994tx} based on unitarity.  However, such constraints are only generally relevant  for spin $>2$.
}

Moving one step further into the complexity,  a higher-dimensional operator can be obtained by coupling to the  Riemann tensor. In the linearized theory, the Riemann tensor is
\begin{equation}
    R_{\mu\nu \rho\sigma }=\frac{\kappa}{2}\left( {h
  }_{\mu\sigma,\nu\rho}+{h
  }_{\nu\rho,\mu\sigma}- {h
  }_{\nu\sigma,\mu\rho}- {h
  }_{\mu\rho,\nu\sigma}\right)\,,
\end{equation}
which in general corresponds to the quadrupole coupling. Such a dimension 6 operator results from  the ${\hat g}_2^{(+2)}$ term:
\begin{equation}
{\hat g}_2^{(+2)}\neq 0,\quad{\hat g}_0^{(+2)} ={\hat g}_1^{(+2)} ={\hat g}_3^{(+2)} =0\quad \longrightarrow\quad    R_{\mu\nu\rho\sigma}\psib^\mu \gamma^\rho\gamma^\sigma\psi^\nu\,.
\end{equation}
The ${\hat g}_3^{(+2)}$ term stems from higher-order multipoles, and requires at least four derivative insertions. We will not discuss such an operator further.

Finally, let us focus on the high-energy limit of the ${\hat g}_0^{(+2)}$ (minimal coupling) and ${\hat g}^{(+2)}_1$  terms. For a positive helicity graviton the ${\hat g}_0^{(+2)}$ coupling becomes
\begin{equation}
\left<\btwo\bthree\right>\left(\left<\btwo\varepsilon^+_1\bthree\right]+\left<\bthree\varepsilon^+_1\btwo\right]\right)^2\rightarrow\left\{ \begin{aligned}
    &4\frac{[12]^5}{[13][23]^2}m_{3/2}\sim\mathcal{O}(E^2)\quad\text{for }\left(+2,+\frac{3}{2},-\frac{3}{2}\right)\\&-4\frac{[13]^5}{[23]^2[12]}m_{3/2}\sim\mathcal{O}(E^2)\quad\text{for }\left(+2,-\frac{3}{2},+\frac{3}{2}\right)\\&-2\frac{[12]^3[13]}{[23]^2}m_{3/2}\sim\mathcal{O}(E^2)\quad\text{for }\left(+2,+\frac{1}{2},-\frac{1}{2}\right)\\&2\frac{[13]^3[12]}{[23]^2}m_{3/2}\sim\mathcal{O}(E^2)\quad\text{for }\left(+2,-\frac{1}{2},+\frac{1}{2}\right)\end{aligned}\right.\label{masslesspsipsih}
\end{equation}
while for the ${\hat g}_1^{(+2)}$ coupling we find
\begin{equation}
\left[\btwo\bthree\right]\left(\left<\btwo\varepsilon^+_1\bthree\right]+\left<\bthree\varepsilon^+_1\btwo\right]\right)^2\rightarrow\left\{ \begin{aligned}
    &\sqrt{\frac{2}{ 3}}\frac{[13]^4}{ [23]} \sim\mathcal{O}(E^3)\quad\text{for }\left(+2,-\frac{1}{2},+\frac{3}{2}\right)\\&-\frac{2\sqrt{2}}{3}\frac{[13]^2[12]^2}{ [23]} \sim\mathcal{O}(E^3)\quad\text{for }\left(+2,+\frac{1}{2},+\frac{1}{2}\right)\\&\sqrt{\frac{2} {3}}\frac{[12]^4}{ [23]} \sim\mathcal{O}(E^3)\quad\text{for }\left(+2,+\frac{3}{2},-\frac{1}{2}\right)
    \end{aligned}\right.
    \label{masslessnonmincoup}
\end{equation}     
where the ${\hat g}_0^{(+2)}$ and ${\hat g}_1^{(+2)}$ terms have mass dimension $-2$. The high-energy limit with the negative helicity graviton can be obtained analogously.

\end{itemize}

\subsection{$(s,1/2, 3/2)$}
\label{sec:s2s3}

We next discuss the three-point interactions $(s,1/2, 3/2)$ with  a single massive spin-3/2 particle, a spin 1/2  and a third particle of spin $s=0,1,2$. 
For the one massless, two massive amplitude, the basis vectors are chosen to be $v_\alpha=\lambda_{1\alpha} $, $u_\alpha=\frac{p_{3\alpha\dotbeta}}{m_1}\Tilde{\lambda}_1^{\dotbeta}$, assuming particle label $1$ is the massless particle, and the amplitude with one spin 3/2 and one spin 1/2 is written as
\begin{equation}
    M^h_{\alpha_1\alpha_2\alpha_3,\beta}= f_1\left(u^{2+h}v^{2-h}\right)^{(1)}_{\alpha_1\alpha_2\alpha_3,\beta}+f_2\left(u^{2+h}v^{2-h}\right)^{(2)}_{\alpha_1\alpha_2\alpha_3,\beta}\,,
    \label{differentmass}
\end{equation}
where $f_{1,2}$ are coupling constants and $(1),(2)$ label the two possible tensor structures from the product $u^{2+h}v^{2-h}$. The results for $s=0,1,2$ are as follows:

\begin{itemize} 
\item $s=0$

    Both particles $2,3$ are massive in our setup, and the spinors at hand are $\left|\bthree\right)$, $\left|\bthree\right)$, $\left|\bthree\right)$, $\left|\btwo\right)$. Without momentum insertion, one cannot write a  three-point amplitude using these spinors --  since little-group indices are symmetrized, the contraction $\left(\bthree\bthree\right)$ is zero. The next possibilities are dimension $5$ operators  with one momentum insertion:
    \begin{equation}\begin{aligned}
    &\text{Parity odd:}\quad   \left<\bthree\right|p_2\left|\bthree\right] \left(\left<\bthree\btwo\right>-       \left[\bthree\btwo\right]\right) \quad \rightarrow\quad i \varepsilon_{\mu\nu\alpha\beta}\psib^\mu \gamma^{\nu\alpha}\chi\partial^\beta P+\text{h.c.}
        \\&\text{Parity even:}\quad  \left<\bthree\right|p_2\left|\bthree\right]        \left(    \left<\bthree\btwo\right>+\left[\bthree\btwo\right]\right) \quad \rightarrow\quad     i  \psib ^\mu \partial_\mu \chi S +\text{h.c.} 
      \label{dim5scalarfermion}
    \end{aligned}\end{equation}
 
    Note that inserting either $p_1$ or $p_2$ is equivalent, due to momentum conservation, or integration by parts from the perspective of the Lagrangian. One may try with  two momentum insertions (for example the dimension 6 operator ${\psib}^\mu\slashed{\partial} \chi \partial_\mu\phi$):
\begin{equation}
        \left<\bthree\right|p_1\left|\bthree\right]        \left<\bthree\right|p_1\left|\btwo\right],\quad \left<\bthree\right|p_1\left|\bthree\right]        \left<\btwo\right|p_1\left|\bthree\right],
    \end{equation}
but these term reduces to \eqref{dim5scalarfermion} upon using momentum conservation and the Dirac equation. Therefore, the three-point amplitude composed of spin $(0,1/2,3/2)$ is always determined by a  linear combination of the amplitudes in \eqref{dim5scalarfermion}.  This is also true when the scalar is massless.

The on-shell formalism also provides a simple way to infer universal properties of spin-3/2  decays, bypassing the need for a Lagrangian -- no matter the dimension of the  interaction, the decay rate can be uniquely specified by the 
number of parity-even/odd contributions  (Eq.~\eqref{dim5scalarfermion}). In addition, for a given particle content, the on-shell approach also pins down the spin-3/2 particle lifetime by combining the decay rates, which again is parametrized only by the parity even/odd contributions. Interesting phenomenological applications can be found, for example, in Ref.~\cite{Bertuzzo:2023slg} for axion-like particles.
 
\item $s=1$ (massless)

For a helicity $+1$ gauge boson, we apply Eq.~\eqref{differentmass} contracted with  massive spinors to give 
\begin{equation}
\begin{aligned}
     M^+&=f_1^{(+1)}  [\bthree 1]^2\left(m_\chi [\btwo\bthree ]-m_{3/2} \left<\btwo\bthree\right>\right)+f_2^{(+1)}\frac{m_\chi}{m_{3/2}}[\bthree 1]^2\left(m_{3/2} [\bthree \btwo]-m_\chi \left<\btwo\bthree\right>\right)\,,
     \\&=[\bthree1]^2 \left({\hat f}_1^{(+1)} [\bthree\btwo]+{\hat f}_2^{(+1)}\left<\bthree\btwo\right>\right)\,,
\end{aligned}
\end{equation} 
and similarly for helicity $-1$ we obtain
\begin{equation}
    M^-= \left<\bthree1\right>^2 \left({\hat f}_1^{(-1)} [\bthree\btwo]+{\hat f}_2^{(-1)}\left<\bthree\btwo\right>\right)\,,
\end{equation}
where we have defined 
\begin{equation}
    {\hat f}_1^{(\pm 1)}  \equiv m_\chi\left(-{ f}_1^{(\pm 1)} +{ f}_2^{(\pm 1)}\right)  ,\quad {\hat f}_2^{(\pm 1)}\equiv  m_{3/2}\left({ f}_1^{(\pm 1)}+\frac{m_\chi^2}{m_{3/2}^2}{ f}_2^{(\pm 1)} \right)\,.
\end{equation}
Interestingly, the dimension 4 operators corresponding to $\left<\bthree \varepsilon_1^+ \bthree\right]\left<\bthree\btwo\right>,   \left<\bthree\varepsilon_1^+ \bthree\right]\left[\bthree\btwo\right]$ are absent, which means that couplings of the type $\psib^\mu \chi A_\mu+\text{h.c.}$ are either forbidden, or they require other interactions to cancel them on-shell, for example $ \psib^\mu(i\slashed{\partial}-m_\chi) \chi A_\mu+\text{h.c.}$. On the other hand, the  
on-shell operators from the amplitudes $[\bthree 1]^2\left<\bthree\btwo\right>$  and $ \left<\bthree1\right>^2\left[\bthree\btwo\right]$  must involve a derivative acting on the gauge boson (in  order to obtain $|3]|3]$ or $\left|3\right>\left|3\right>$ from the polarization vector). A  dimension $5$ example is the     coupling of the gauge boson field strength to spin 1/2 and spin 3/2:
\begin{equation}\text{Parity even:}\quad [\bthree1]^2\left<\bthree\btwo\right>,\left<\bthree1\right>^2\left[\bthree\btwo\right]
\quad 
\longrightarrow\quad i \bar{\chi}\gamma^\mu\psi^\nu F_{\mu\nu}+\text{h.c.}  
\label{eq:evenpdim5}
\end{equation}
Another vertex equivalent to \eqref{eq:evenpdim5} is  $i\bar{\chi}\gamma^\rho\gamma^{\mu\nu}\psi_\rho F_{\mu\nu} +\text{h.c.}$, which features in the supergravity Lagrangian. The parity odd interaction is constructed from  the dual field strength $\Tilde{F}_{\mu\nu}$
\begin{equation}
\text{Parity odd:}\quad [\bthree 1]^2\left<\bthree\btwo\right>,-\left<\bthree1\right>^2\left[\bthree\btwo\right]
\quad 
\longrightarrow\quad i \bar{\chi}\gamma^\mu\psi^\nu \Tilde{F}_{\mu\nu}+\text{h.c.}  \label{odddim5}
\end{equation}
The ${\hat f}_1^{(+1)}$, ${\hat f}_2^{(-1)}$ terms in the basis require at least two derivatives. Examples of dimension $6$ operators are
\begin{equation}\begin{aligned}
    \text{Parity even:}\quad  &\left[\bthree\btwo\right]\left[\bthree1\right]^2- \frac{m_\chi}{m_{3/2}}\left<\bthree\btwo\right>\left[\bthree1\right]^2, \left<\bthree\btwo\right>\left<\bthree1\right>^2- \frac{m_\chi}{m_{3/2}}\left[\bthree\btwo\right]\left<\bthree1\right>^2 \\&\longrightarrow\quad   \psib^\mu  \gamma^{\nu\rho }\chi \partial_\mu F_{\nu\rho}+\text{h.c.}
\end{aligned}
\end{equation}
and analogously for the parity odd operator
\begin{equation}\begin{aligned}
    \text{Parity odd:}\quad & \left[\bthree\btwo\right]\left[\bthree1\right]^2- \frac{m_\chi}{m_{3/2}}\left<\bthree\btwo\right>\left[\bthree1\right]^2, -\left<\bthree\btwo\right>\left<\bthree1\right>^2+ \frac{m_\chi}{m_{3/2}}\left[\bthree\btwo\right]\left<\bthree1\right>^2 \\&\longrightarrow\quad   \psib^\mu  \gamma^{\nu\rho }\chi \partial_\mu \Tilde{F}_{\nu\rho}+\text{h.c.}
\end{aligned}
\end{equation}

\item $s=2$ 

There is only one structure according to 
Eq.~\eqref{differentmass}, namely  
\begin{equation}
M^{+2}=f_1^{(+2)}[\bthree1 ]^3[\btwo1],\qquad M^ {-2}=f_1 ^{(-2)}\left<\bthree1\right>^3\left<\btwo1\right>\,.
\end{equation}
They are obtained with at least 3 derivatives, namely dimension $\geq7$ operators, that can take the form, for example,~$R_{\mu\nu\alpha\beta} \partial^\mu\psib^\nu \gamma^{\alpha\beta}\chi$. We will not be interested in such high-dimensional operators in this work.

 \end{itemize}

\subsection{Implications from IR unification}\label{gaugeinvsec}

So far we have obtained a collection of three-point amplitudes consistent with little-group covariance, describing interactions   of the spin-3/2 particle with other particles of spin $\leq 2$. This basis is fully general, and the only model dependence is the particle content, which can be easily modified or extended. 
In section~\ref{secfourpoint}, we will discuss how higher-point amplitudes are constructed from lower point amplitudes using recursion relations. However, before proceeding, 
a crucial remark is that such a construction is not possible for an arbitrary theory and therefore, one must first specify the three-point interactions that provide the basic building blocks.

Since we will consider spin-3/2 scattering in section~\ref{secfourpoint}, we focus on vertices from section~\ref{s3232} containing two spin-3/2 particles and interactions of dimension $\leq 5$. While the superHiggs mechanism will be recovered on-shell from the bottom-up in the next section and couplings can be fixed by perturbative unitarity, some constraints also arise from a top-down point of view.  
In the on-shell language, gauge symmetry does not get ``broken'' at low energy, but rather, in the IR, massless amplitudes with different helicities unify into a single massive amplitude, where the mass appears as a pole $1/m$~\cite{Arkani-Hamed:2017jhn}, due to the polarization vector~\eqref{eq:massivepolvect}. 
Reversing this logic then implies that the $1/m$ singularity in the massive amplitude must disappear in the high-energy limit. At the three-point level, the requirement of a smooth massless limit therefore imposes non-trivial conditions on the couplings.  

To start with, the two spin-3/2 polarization vectors include the mass pole $1/ m_{3/2}^2$, which then  must disappear in the massless (high-energy) limit. For the scalar coupling $ \psib^{ \mu}\psi _\mu  S$ in \eqref{phipsipsiinteraction}, the three-point amplitude reads
\begin{equation}
    \bar{v}^{\mu}u_\mu =\frac{1}{m_{3/2}^2}\left(\left< \btwo\bthree\right>^2\left[\btwo\bthree\right]+\left<\btwo\bthree\right>\left[\btwo\bthree\right]^2\right)\,,
\end{equation} 
where $\psib^\mu \propto {\bar v}^\mu$, $\psi_\mu \propto u_\mu$ using \eqref{eq:polvec32}. 
The high-energy limit of the three-point amplitudes Eq.~\eqref{masslesspsipsis1}-\eqref{masslesspsipsis2} then implies that $\psib^\mu \psi_\mu S \sim 1/m_{3/2}$, which still has a mass pole. In order to cancel this pole, the coupling constant should therefore be proportional to $m_{3/2}$ and by dimensional analysis, this results in the low-energy coupling $\propto \kappa m_{3/2}\psib^\mu \psi_\mu S $.
 The same argument carries over to the pseudoscalar coupling $i\varepsilon^{\mu\nu\alpha\beta } \partial_\mu P \psib_\nu \gamma_\alpha\psi_\beta $ in \eqref{phipsipsiinteraction}, which gives on-shell the three-point amplitude
\begin{equation} -\varepsilon^{\mu\nu\alpha\beta } p_{1\mu}  \bar{v}_\nu  \gamma_\alpha u_\beta =-\frac{2}{m_{3/2}}  i \left(\left< \btwo\bthree\right>^2\left[\btwo\bthree\right]-\left<\btwo\bthree\right>\left[\btwo\bthree\right]^2\right)\,.
\label{eq:pseudoonshell}
\end{equation} 
Note that due to the derivative coupling, the equation of motion cancels one $1/m_{3/2}$ factor. 
Again using Eq.\,\eqref{masslesspsipsis1}-\eqref{masslesspsipsis2}, a smooth high-energy limit is then guaranteed. Since \eqref{eq:pseudoonshell} is a dimension 5 interaction, the coupling constant is proportional to $\kappa$, namely we obtain a coupling $\propto i\kappa\varepsilon^{\mu\nu\alpha\beta } \partial_\mu P \psib_\nu \gamma_\alpha\psi_\beta $.

Finally, for the spin-3/2 coupling to the graviton, the two dimension $5$ operators correspond to the ${\hat g}_0^{(+2)}$, ${\hat g}_1^{(+2)}$ terms in the expansion \eqref{gravitonplus}, with  high-energy limits \eqref{masslesspsipsih}, \eqref{masslessnonmincoup}, respectively. Given that the coupling constant is proportional to $\kappa$ and the massive amplitude has a $1/m_{3/2}$ pole\,\footnote{Of course, it is always possible to replace $\kappa$ by $m_{3/2}^n\kappa^{n+1}$ ($n\geq 1$) in the coupling coefficients in order to cancel the arbitrary mass pole, but such a coupling is much more Planck-suppressed, and therefore will not be interesting for us.} (see e.g. \eqref{mincoupinspinor}), only the minimal coupling ${\hat g}_0^{(+2)} $ has a smooth massless limit. Thus, very interestingly, the constraint from IR unification points to the gravitational minimal coupling, $\propto\kappa h_{\mu\nu}T^{\mu\nu}_{3/2}$. 

In summary, imposing the ``IR unification" of massless amplitudes into massive amplitudes, as advocated in Ref.~\cite{Arkani-Hamed:2017jhn}, we obtain the following vertices involving two massive spin-3/2 particles 
 \begin{equation}
    \begin{aligned} c_S\, m_{3/2}  \kappa\, \psib^\mu\psi_\mu S+ c_P\, i\kappa \varepsilon^{\mu\nu\alpha\beta } \partial_\mu P \psib_\nu \gamma_\alpha\psi_\beta  + c_h\kappa \,h_{\mu\nu }T^{\mu\nu}_{3/2}\,,
     \end{aligned}
\label{psipsicouplings}
\end{equation}
where $c_S, c_P$ and $c_h$ are arbitrary  dimensionless coefficients.

Similarly, we can check the high-energy limit of the  dimension $\leq 5$ operators in section~\ref{sec:s2s3} involving just one spin-3/2 particle, that may be useful in, for example, gravitino-chiral fermion scattering, or Compton scattering with spin 1/2 exchange. In fact, all operators in Eq.~\eqref{dim5scalarfermion}, \eqref{eq:evenpdim5}-\eqref{odddim5} have a smooth massless limit.\footnote{More precisely, while taking the massless limit, the coefficient of the three-point amplitude is either $m_{3/2}/m_{3/2}=1$, which means that the mass pole cancels, or $m_\chi/m_{3/2}$. For the latter, if $\chi$ is considered to be the longitudinal component of the massive spin 3/2, its mass is then $m_\chi=m_{3/2}$, which also  cancels the pole. For $m_\chi\neq m_{3/2}$, a smooth massless limit is possible if $m_\chi\rightarrow 0$ faster than $m_{3/2}\rightarrow 0$. 
} 
Thus, for completeness, we present the massive amplitudes below with dimensionless coupling constants $a_P$, $a_S$, $a_A$, $\tilde{a}_A$: 
\begin{equation}
\begin{aligned}
  &  i a_P\kappa \varepsilon_{\mu\nu\alpha\beta}\psib^\mu \gamma^{\nu\alpha}\chi\partial^\beta P+\text{h.c.}, \qquad  i a_S \kappa \psib ^\mu \partial_\mu \chi S +\text{h.c.} ,\\& i a_A \kappa \bar{\chi}\gamma^\mu\psi^\nu  F_{\mu\nu}+\text{h.c.}  ,\qquad i \tilde{a}_A\kappa \bar{\chi}\gamma^\mu\psi^\nu \Tilde{F}_{\mu\nu}+\text{h.c.}  
\end{aligned}
\label{eq:psicouplings}
\end{equation}
The high-energy limit of the couplings in \eqref{eq:psicouplings} give rise to the massless three-point amplitudes with helicity $(\pm3/2,\pm1/2,0)$ (scalar coupling -- also given in Ref.~\cite{Arkani-Hamed:2017jhn}), and $(\pm3/2,\mp1/2,\pm1)$, $(\pm 1/2,\pm 1/2, \pm1)$ (gauge boson coupling).  

Finally, we mention an alternative bottom-up approach 
that also determines the three-point interactions with a smooth massless limit. While studying higher-spin Compton amplitudes, it has been noticed in \cite{Cucchieri:1994tx} that a necessary condition to have a unitary theory is that 
$\partial^\mu J_\mu =\mathcal{O}(m)$ where the current $J_\mu$ associated with a particle $\phi$ of spin $\geq1$ and mass $m$ is defined in \eqref{eq:currentdefn}.
Applied to spin-3/2 particles, this condition 
also amounts to a ``smooth  massless limit'' for the current, but the perspective here is different, because the field associated with the current  is off-shell. In the end, imposing this current condition also gives rise to the interactions in \eqref{psipsicouplings} and \eqref{eq:psicouplings}. For further details on this point of view, see appendix~\ref{app:gaugeinv}.

\section{Four-point amplitudes and unitarity}
\label{secfourpoint}

Equipped with the on-shell three-point interactions \eqref{psipsicouplings} for spin 3/2 derived in section~\ref{gaugeinvsec}, we will next construct four-point amplitudes. For simplicity, we will restrict the particle content to one massless spin 2, one massive spin 3/2, one massive scalar and one massive pseudoscalar. This setup can be easily generalized. In addition, we only consider interactions with dimension $\leq5$.

The on-shell formalism allows the construction of  higher-point amplitudes from lower-point building blocks, thanks to recursion relations. In the massless case, this formalism is known to be particularly powerful, such as in computing amplitudes with $n$   gluons or gravitons  \cite{Berends:1987me,Bern:2007dw,Dixon:2010ik,Bedford:2005yy,Cachazo:2005ca}. For massive amplitudes, the construction can be achieved by finding an appropriate momentum shift. This momentum shift is crucial in the on-shell construction because simply ``gluing'' together three-point amplitudes without shifting the momentum will cause a contact term ambiguity in the resulting four-point amplitude.
In the following, we use the  ``all-line transverse" shift proposed in \cite{Ema:2024rss,Ema:2024vww} to construct the tree-level gravitino scattering amplitude.

\subsection{Momentum shift and on-shell recursion relation}

We begin by briefly reviewing the on-shell recursion relations as well as defining the all-line transverse momentum shift.  
The on-shell construction of higher-point amplitudes is based on the deformation of external momenta in the complex plane, with a complex parameter $z$ and constant four-vector $q_i$, given by
\begin{equation}
    \hat{p}_i=p_i+ z q_i\,.
\end{equation}
Under this shift, the resulting amplitude is a meromorphic function of $z$. In general, we require that the shifted momenta satisfy the on-shell condition and momentum  conservation, 
namely,
\begin{equation}\hat{p}_i^2=m_i^2,\qquad 
    \sum_i \hat{p}_i=0\,.
\end{equation}

At tree level, locality ensures that the amplitude has poles when the internal momenta $p_I$ are on shell. The Cauchy theorem in complex analysis gives:
\begin{equation}
\begin{aligned}
	 {A}_n &= \frac{1}{2\pi i}\oint_{z=0} \frac{dz}{z} \hat{A}_n(z)
  = - \sum_{\{z_I\}} \text{Res}\left[\dfrac{\hat{A}_n(z)}{z}\right] + B_\infty\,,
  \label{eq:CauchyAn}
\end{aligned}
\end{equation}
where $z_I$ are the solutions of the pole condition $\hat{p}^2_I(z_I)=m_I^2$, with $m_I$ the mass of the internal propagating particle. The hatted amplitude $\hat{A}_n$ in \eqref{eq:CauchyAn} depends on the shifted momenta and the constant $B_\infty$ corresponds to the boundary term at $|z|\rightarrow \infty$. The pole structure of amplitudes  then dictates that  $A_n$ at the poles factorizes into a product of subamplitudes, multiplied by a propagator~\cite{Weinberg_1995}
\begin{equation}
    \begin{aligned}
	A_n = -\sum_{z = z_I}\sum_\lambda\mathrm{Res}\left[\hat{A}^{(\lambda)}_{n-m+2}\frac{1}{z}\frac{1}{\hat{p}_I^2 - m_I^2}
	\hat{A}^{(-\lambda)}_{m}\right]
	+ B_\infty,
\end{aligned}\label{npointconstruction}
\end{equation}
where $\lambda$ denotes the helicity of the intermediate state. 

Note that the construction of higher-point amplitudes  
crucially depends on the behavior of $\hat{A}(z)$ at $|z|\rightarrow \infty$, i.e.~the boundary term $B_\infty$.  If $B_\infty$ is zero for a certain momentum shift, we say the theory has a ``good'' large-$z$ behavior, in this case the scattering amplitudes can be constructed solely from lower point subamplitudes. In particular, for four-point amplitudes we obtain
\begin{equation}
\begin{aligned}
	A_4 = -\sum_{z = z_I}\sum_\lambda\mathrm{Res}\left[\hat{A}^{(\lambda)}_{3}\frac{1}{z}\frac{1}{\hat{p}_I^2 - m_I^2}
	\hat{A}^{(-\lambda)}_{3}\right]\,,
\end{aligned}\label{four-point-construction}
\end{equation}
assuming the boundary term vanishes.  The sum in \eqref{four-point-construction} covers all possible channels and intermediate particles. Note that in \eqref{four-point-construction}, the three-point amplitudes are shifted by  $z$-dependent quantities, and must be evaluated at the poles  $z=z_I$. 

In summary, to construct higher-point amplitudes based on the analytic continuation in the complex plane, 
one cannot simply ``glue'' the on-shell three-point amplitudes without (1) shifting the momenta, (2) discussing the large-$z$ behavior of the deformed amplitude, in other words the boundary term $B_\infty$, and (3) evaluating the subamplitudes at the poles.

Whether or not a certain momentum shift leads to a good large-$z$ behavior depends on the mass, spin and helicity of the particles, as well as the type of interaction.  
In the massless  case, the BCFW recursion relation \cite{Britto:2004ap,Britto:2005fq} is frequently used. It consists of shifting two of the four external momenta
\begin{equation}
    \hat{p}_1=p_1+ z q,  \quad \hat{p}_3=p_3- z q\,,
\end{equation}
where the constant momentum $q$ is chosen such that $p_{1,3}\cdot q= q^2=0$, so both momentum conservation and the on-shell conditions are satisfied.

On the other hand, the construction of massive amplitudes has been subject to contact term ambiguities \cite{Christensen:2022nja}, and  without an appropriate deformation of the momentum, the result does not match that obtained from Feynman rules. Recently, a momentum shift for massive amplitudes, called the \textit{all-line  transverse} (ALT) shift, was  proposed in \cite{Ema:2024vww,Ema:2024rss} and shown to be appropriate for renormalizable theories with spin $\leq1$, including QED and the electroweak theory. In this case, all external momenta are shifted by a transverse polarization vector, hence the name ALT shift. Further details about the ALT shift for spin $\leq1$, and the argument for $B_\infty=0$, are reviewed in appendix \ref{app:recursion}.

\subsection{All-line transverse shift for spin 3/2}
\label{sec:alt32}

Next, we extend our discussion about constructibility to massive spin-3/2 particles. Most of the arguments for spin 1/2 and spin 1 (see appendix \ref{app:recursion}) still apply and the definition of the ALT shift carries over to the spin-3/2 transverse states. As we show below, the transverse polarization vectors remain unchanged, whereas the longitudinal ones depend linearly on $z$ under this shift.
The polarization vectors are given in Eq.~\eqref{gravitino-helicities}, which, in terms of the helicity basis, can be rewritten as
\begin{equation}
    \begin{aligned}
     &   \psi^{++}=\frac{\sqrt{2}}{m_{3/2}}\left(\begin{array}{c}\Tilde{\lambda}_{\dotalpha}\eta_\alpha\eta_\beta  \\
    \Tilde{\lambda}_{\dotalpha} \eta_\alpha\Tilde{\lambda}^{\dotbeta}
        \end{array}\right),\quad \psi^{+}=\frac{\sqrt{2/3}}{m_{3/2}}\left(\begin{array}{c}\lambda_\alpha            \Tilde{\lambda}_{\dotalpha}\eta_\beta 
 +\lambda_\beta               \Tilde{\lambda}_{\dotalpha}\eta_\alpha +\eta_\alpha\eta_\beta\tilde{\eta}_{\dotalpha}
 \\\eta_\alpha               \Tilde{\eta}_{\dotalpha}\Tilde{\lambda}^{\dotbeta}+\eta_\alpha               \Tilde{\eta}^{\dotbeta} \Tilde{\lambda}_{\dotalpha}+\tilde{\lambda}_{\dotalpha}\tilde{\lambda}^{\dotbeta}\lambda_\alpha
        \end{array}\right)\,,\\&  
        \psi^{--}=\frac{\sqrt{2}}{m_{3/2}}\left(\begin{array}{c}
               \tilde{\eta}_{\dotalpha}\lambda_\alpha\lambda_\beta  \\               \tilde{\eta}_{\dotalpha}\lambda_\alpha\Tilde{\eta}^{\dotbeta}
        \end{array}\right),\quad      \psi^{-}=\frac{\sqrt{2/3}}{m_{3/2}}\left(\begin{array}{c}\lambda_\alpha               \Tilde{\lambda}_{\dotalpha}\lambda_\beta 
 +\eta_\beta                {\lambda}_{\alpha}\tilde{\eta}_{\dotalpha} +\lambda_\alpha\eta_\beta\tilde{\eta}_{\dotalpha}
 \\\lambda_\alpha               \Tilde{\lambda}_{\dotalpha}\Tilde{\eta}^{\dotbeta}+\lambda_\alpha               \Tilde{\lambda}^{\dotbeta} \Tilde{\eta}_{\dotalpha}+\tilde{\eta}_{\dotalpha}\tilde{\eta}^{\dotbeta}\eta_\alpha
        \end{array}\right)\,.
    \end{aligned}
\end{equation}
Similarly to the spin 1/2 and spin 1 cases, the momentum shift for helicity $+3/2$ and $-3/2$ is respectively \eqref{positiveshift} and \eqref{negativeshift}, which leaves the polarization vectors $\psi^{++}$, $\psi^{--}$ unchanged. 

However, contrary to spin 1,  the longitudinal helicities $\pm 1/2$ will inevitably depend on $z$, because they are an admixture of $\varepsilon^{0,\pm}$ and $u^\pm$. We may attempt to shift $\psi^+$ by \eqref{positiveshift} and $\psi^-$ by \eqref{negativeshift}\footnote{If we shift $\psi^\pm$ in the opposite way, there will be extra terms that are quadratic in the shift parameters $w,{\tilde w}$.} to give  
\begin{equation}\begin{aligned}
  &  \psi^+\rightarrow \psi^+ +\frac{1}{m_{3/2}}\sqrt{\frac{2}{3}}\left(\begin{array}{c}
             (2w+\tilde{w}) \Tilde{\lambda}_{\dotalpha}\eta_\alpha\eta_\beta  \\ (w+2\tilde{w})
               \Tilde{\lambda}_{\dotalpha} \eta_\alpha\Tilde{\lambda}^{\dotbeta}
        \end{array}\right)= \psi^+ +\frac{1}{\sqrt{3}} \left(\begin{array}{cc}2w+\tilde{w}& 0\\0 &w+2\tilde{w}  \end{array}\right)\psi^{++}\,,
        \\&
        \psi^-\rightarrow \psi^- +\frac{1}{m_{3/2}}\sqrt{\frac{2}{3}}\left(\begin{array}{c} (w+2\tilde{w}) \tilde{\eta}_{\dotalpha}\lambda_\alpha\lambda_\beta  \\ (2w+\tilde{w})              \tilde{\eta}_{\dotalpha}\lambda_\alpha\Tilde{\eta}^{\dotbeta}
        \end{array}\right)= \psi^- +\frac{1}{\sqrt{3}} \left(\begin{array}{cc}w+2\tilde{w}& 0\\0 &2w+\tilde{w}  \end{array}\right)\psi^{--}\,,
\end{aligned}
\end{equation}
but the resulting $\psi^\pm$ cannot be invariant for any nonzero $w,\tilde{w}$, which are independent shift parameters. In consequence, the $z$-dependent  polarization vectors will worsen the large-$z$ behavior of the amplitude, eventually giving rise to boundary terms.  

For now, we assume that all external spin-3/2 states are transverse, so that   the $z$-dependence originates only from the propagator and the derivatives. One can carry out  the dimensional analysis in appendix \ref{app:recursion} to determine the constructibility. Since the discussion depends on the specific theory and the amplitude, we will illustrate the idea in several examples in the following. In certain cases, the Ward identity must be \textit{imposed}, i.e.~we pin down the theory to be that satisfying the Ward identity, in order to have constructibility.

\vspace{0.5cm}

\textbf{\textit{Four-point spin-3/2 amplitude:}} 
To construct a four-point spin-3/2 amplitude, the relevant three-point vertices must involve two spin-3/2 states. If the interaction term has dimension $\geq6$, the derivative couplings in each operator contribute with $z^{n\geq 2}$, while the (bosonic) propagator\footnote{In the case where the intermediate particle is e.g.~a massive spin 1, then the propagator has a $p_\mu p_\nu/m^2$ piece in the numerator, which a priori brings an additional $z^2$ dependence at large $z$. Such  contribution can be avoided if one imposes smooth massless limit of the theory, see section \ref{gaugeinvsec} or appendix \ref{app:gaugeinv}.} scales as $z^{-2}$ for large $z$,  so the total large-$z$ growth is faster than $z^{2}$. Even after imposing the Ward identity, the boundary term does not vanish.

Restricting now to dimension $\leq5$ operators,  the scattering amplitude has at most two derivatives, and after the ALT shift the $z$ dependence of the derivatives (momenta) can be as large as $z^2$. Taking into account the propagator, the total large-$z$ behavior can be as large as $z^{0}$ which naively does not vanish at complex infinity. Next, we will argue that the boundary term is absent when the Ward identity is imposed, similarly to the previous subsection.

The amplitude can be decomposed by factoring out one of the transverse polarization vectors:
\begin{equation}
    A_4=\psi_{i\mu}^{\pm\pm} F^\mu(p_1,p_2,p_3,p_4)\,,
\end{equation}
where $F^\mu$ represents the matrix element with one external polarization stripped off. At large $z$, the shifted amplitude is dominated by the amplitude with all momenta replaced by the shifted momenta $z  r_i$, namely
\begin{equation}
 \lim_{z\rightarrow\infty} \hat{A}_4 = z^{n\leq 0} \psi_{i\mu}^{\pm\pm} F^\mu(r_1,r_2, r_3,r_4)\propto  z^{n\leq 0} u_{i}^{\pm}r_{i\mu} F^\mu(r_1,r_2, r_3,r_4)\,.
 \label{highestzamp}
\end{equation}
The last step is due to the property that $r_i\propto \varepsilon^{\pm}_i$. 
 Then $F_\mu(r_i)$ encodes the massless limit of the spin-3/2 scattering where each   $\psi_i$ is associated with a light-like momentum $r_i$. We can easily check that the massless equation of motion and constraints are fulfilled:
 \begin{equation}
 \begin{aligned} & \text{Gamma trace constraint:}\quad \gamma^\mu\psi_{i\mu} ^{\pm\pm}=0\,.
\\ &  \text{Divergence constraint:}\quad r_i^\mu\psi_{i\mu} ^{\pm\pm}\propto \varepsilon_{i\mu}^{\pm}\varepsilon_{i}^{\mu\,\pm}u_i^{\pm}=0\,.\\
&  \text{Equation of motion:}\quad \slashed{r}_i \psi_{i\mu} ^{\pm\pm}\propto \gamma^\nu \varepsilon_{i\nu}^{\pm}  \varepsilon_{i}^{\mu\,\pm}u_i^{\pm}\propto  \varepsilon_{i}^{\mu\,\pm} \gamma^\nu\psi_{i\nu} ^{\pm\pm}=0\,.
\end{aligned}
\label{eq:constraints}
\end{equation}

If one imposes the Ward identity, then $r_{i\mu } F^\mu(r_i)=0$ so $\lim_{z\rightarrow\infty} \hat{A}_4 = 0$, hence the deformed amplitude gives rise to no boundary term at $|z|\rightarrow\infty$. In other words, this four-point amplitude is constructible under the ALT shift. 
In section~\ref{fourpointsec}, we will  explicitly compute this four-point amplitude and derive the contact term, using the ALT shift.\\

\textbf{\textit{n-point spin-3/2 amplitude:}}  
We next consider the constructibility for $n\geq 5$, assuming
all external particles are spin-3/2 transverse states, and   all operators have dimension $\leq5$.  The product of all the couplings then has dimension $[g]\geq-n+2$.
By employing \eqref{naivegamma} we immediately obtain 
\begin{equation}
\gamma    \leq 4-n-[g]-\frac{n}{2}\leq  2-\frac{n}{2}\,,
\end{equation}
so for $n\geq5$, we have $\gamma<0$ and hence the amplitude is constructible under the ALT shift, namely it can be built from lower point subamplitudes. \\

\textbf{\textit{Compton amplitudes:}}
The (electromagnetic\footnote{The Majorana spin-3/2 particle cannot be charged under electromagnetism, and therefore the electromagnetic Compton scattering through spin 3/2 exchange is not possible in this case. However, the particle content can be extended to include Dirac fermions, which can be charged.} or gravitational) Compton scattering generally has an improved large-$z$ behavior compared to purely massless amplitudes, due to the massless external particles. This can be seen from the photon polarization vector
\begin{equation}
\varepsilon^+_{\alpha\dotalpha}=  \sqrt{2}\frac{\left|\zeta\right>_\alpha[p|_{\dotalpha}}{\left<p\zeta\right>},\quad \varepsilon^-_{\alpha\dotalpha}=  \sqrt{2}\frac{\left|p\right>_\alpha[\zeta|_{\dotalpha}}{\left[p\zeta\right]}\,,
\end{equation}
where $\zeta$ is a reference spinor. For positive (negative) helicity, one can choose to shift $\left|p\right>$ $(\left|p\right])$ by another spinor which is different from $\zeta$, then the denominator of the polarization vector develops a $z$-dependence, and each one contributes with $z^{-1}$ at large-$z$. Analogously, the graviton polarization vector behaves as $z^{-2}$ at complex infinity. As before, we assume that the external massive spin-3/2 states are transverse so that they are unchanged under the ALT shift.

We can put further constraints on the theory to   specify the large-$z$ behavior. Following the analysis in \cite{Cucchieri:1994tx}, tree-level unitarity of Compton amplitudes with higher spins requires that the three-point interaction be gauge invariant in the massless limit (as discussed in appendix~\ref{app:gaugeinv}). Imposing this condition, the $p_\mu p_\nu/m^2$  and $p_{[\mu} \gamma_{\nu]}/m$ terms in the polarization sum of the intermediate spin-3/2 particle will cancel, so the fermionic propagator behaves as $1/z$. In this case, as long as the interaction term has dimension
lower than 5 (6) for electromagnetic (gravitational)  Compton  scattering, the large-$z$ behavior grows no larger than $1/z$, hence the amplitude is constructible. In this case, we do not need to impose the Ward identity associated with the four-point amplitude.  \\

\textbf{\textit{Amplitudes with longitudinal states:}} Our discussion so far, has been restricted to transverse external spin-3/2 states.  We have seen that the longitudinal polarization of spin 3/2 is $z$-dependent under the ALT shift, and therefore one would expect that the longitudinal spin-3/2 scattering is not constructible. Nevertheless, one can exploit the little-group covariance of helicity amplitudes to recover an amplitude with longitudinal states from an amplitude with transverse states only, by applying spin raising or lowering operators. This idea is demonstrated in the spin $1$ case ($WW$ scattering) \cite{Ema:2024rss}. 
Thus, it suffices to prove the constructibility of all transverse  amplitudes to recover, eventually, all the longitudinal amplitudes, even though the latter is a priori not constructible under the ALT shift.

\subsection{Gravitino scattering amplitude}
\label{fourpointsec}

To illustrate the constructibility of four-point amplitudes from the on-shell three-point amplitudes derived in section~\ref{sec:trilinear}, we next consider the scattering of the gravitino (as an example of a spin-3/2 particle) in order to make a connection with known results in supergravity. Assuming the particle content of section~\ref{sec:trilinear} and interaction terms with dimension $\leq 5$, we will construct the gravitino scattering amplitude using the ALT shift and constrain the couplings by perturbative  unitarity. 
The  relevant on-shell three-point interaction terms are given in Eq.~\eqref{psipsicouplings}. 

We will first construct the four-point amplitude $\psi_{\mu}\psi_{\mu}\rightarrow \psi_{\mu}\psi_{\mu}$, with all transverse gravitinos ($h=\pm 3/2$), since as noted in section~\ref{sec:alt32}, the amplitudes involving longitudinal states ($h=\pm 1/2$) can be obtained by applying the spin raising and lowering operators. 
First, we consider the scattering amplitude mediated by the exchange of a graviton. The three-point vertex obtained from 
\eqref{mincouponshell} is 
\begin{equation}
\begin{aligned}
    &V^{h\psi_1\psi_2}(p_1,p_2,p_I)\equiv \varepsilon^{\mu\nu}(p_I)\bar{\psi}_{2}^\alpha \psi_{1}^\beta V_{\mu\nu,\alpha,\beta}^{h\psi_1\psi_2}(p_1,p_2,p_I)\,,
    \\&=c_h\kappa\,\varepsilon^{\mu\nu}(p_I)
    \left[\frac{1}{2}\bar{\psi}_{2\alpha} \gamma_{(\mu}(p_1-p_2)_{\nu)}\psi_{1}^\alpha-\Bar{\psi}_{2\alpha }\gamma_{(\mu}(p_1-p_2)^\alpha\psi_{1\nu)}-\Bar{\psi}_{2(\nu}\gamma_{\mu)}(p_1-p_2)_\alpha\psi_{1}^\alpha\right]\,. 
\end{aligned}
\label{Vhpsipsiexp}
\end{equation} 
Note that all states, including the graviton, are on-shell in this vertex. This feature of the bottom-up, on-shell amplitude approach simplifies the usual Feynman diagram calculation of the four-point amplitude which would use the more complicated off-shell interaction \eqref{gravitinotmunu}. The sum over the graviton polarization vector gives
\begin{equation}
    \sum_{\lambda=\pm2}\varepsilon_\lambda^{*\mu\nu}(p_I)\varepsilon^{\alpha\beta}_\lambda(p_I)=\frac{1}{2}(\hat{\eta}^{\mu\alpha}\hat{\eta}^{\nu\beta}+\hat{\eta}^{\mu\beta}\hat{\eta}^{\nu\alpha}-\hat{\eta}^{\mu\nu}\hat{\eta}^{\alpha\beta})\,,
\end{equation}
where $\hat{\eta}_{\mu\nu}\equiv {\eta}_{\mu\nu} -\frac{p_{I\mu}\bar{p}_{I\nu}+p_{I\nu}\bar{p}_{I\mu}}{p_I\cdot\bar{p}_I}$ and $p_I^\mu=(E,\Vec{p})$ is the graviton momentum with $\Bar{p}_I^\mu\equiv (E,-\Vec{p})$. When the graviton is coupled to the energy-momentum tensor of external particles (up to terms vanishing on-shell), one can use the relation:
 \begin{equation}
     p_{I \mu}T^{\mu\nu}(p_1,p_2)=
     p_{I \nu}T^{\mu\nu}(p_1,p_2)=0\,,\quad\quad p_{I\mu}=p_{1\mu}+p_{2\mu}\,,
 \end{equation}
which amounts to energy conservation. Then we have the simplification
\begin{equation}
  \frac{1}{p_I^2}  T_{\mu\nu}  \sum_{\lambda=\pm2}\varepsilon_\lambda^{*\mu\nu}(p_I)\varepsilon_\lambda^{\alpha\beta}(p_I)T^\prime_{\alpha\beta}= T_{\mu\nu}\Pi^{\mu\nu\alpha\beta}T^\prime_{\alpha\beta}\,,
\label{simpTPIT}
\end{equation}
where $T_{\mu\nu}$, $T^\prime_{\mu\nu}$ can represent two  identical or different spin-3/2 particles and $\Pi^{\mu\nu\alpha\beta}$ is equal to the graviton propagator
\begin{equation}\Pi^{\mu\nu\alpha\beta}=
    \frac{1}{2p_I^2}( {\eta}^{\mu\alpha} {\eta}^{\nu\beta}+ {\eta}^{\mu\beta} {\eta}^{\nu\alpha}- {\eta}^{\mu\nu} {\eta}^{\alpha\beta})\,.
\end{equation}
Note that this is also true \textit{after the shift} because the on-shell condition (for external states) and momentum conservation is always preserved.  

We then proceed to calculate the scattering amplitude $\psi_{\mu}\psi_{\mu}\rightarrow \psi_{\mu}\psi_{\mu}$. There are three factorization channels as shown below, which correspond to internal momenta $p_I=p_1+p_2$, $p_1+p_3$, $p_1+p_4$ going on-shell. They are noted as $s$-, $t$-, $u$-channels, respectively. 
\begin{center}
\begin{tabular}{ccc}
    \adjustbox{valign=m}{\includegraphics[width=0.3\textwidth]{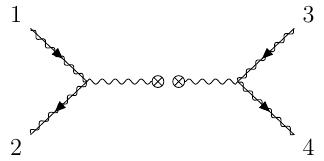}} \hspace{1cm}  & \adjustbox{valign=m}{\includegraphics[width=0.14\textwidth]{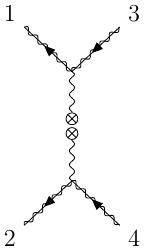}} \hspace{1cm} & \adjustbox{valign=m}{\includegraphics[width=0.14\textwidth]{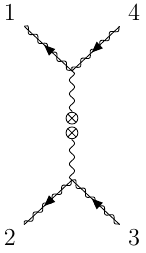}}
    \end{tabular}\end{center}
Starting with the $s$-channel, when all external particles are transverse, under the ALT shift, the vertex \eqref{Vhpsipsiexp} becomes
\begin{equation}
     V^{h\psi_1\psi_2}(\hat{p}_1,\hat{p}_2,\hat{p}_I)=    V^{h\psi_1\psi_2}(p_1,p_2,p_I)+z    V^{h\psi_1\psi_2}(r_1,r_2,r_I)\,.
\end{equation}
The linear $z$ dependence arises from the derivative in the interaction. The four-point amplitude is obtained by multiplying the two shifted subamplitudes together with the shifted propagator  and  then taking the residue. This gives
\begin{equation}\begin{aligned}
     &V^{h\psi_1\psi_2}(\hat{p}_1,\hat{p}_2,\hat{p}_I) V^{h\psi_3\psi_4}(\hat{p}_3,\hat{p}_4,-\hat{p}_I)\\
     &=
     V^{h\psi_1\psi_2}(   p_1, p_2,   p_I) V^{h\psi_3\psi_4}(  p_3,  p_4,-  p_I)+z
     V^{h\psi_1\psi_2}(   p_1, p_2,   p_I) V^{h\psi_3\psi_4}(  r_3,  r_4,-  r_I)
\\&\quad +z
     V^{h\psi_1\psi_2}(   r_1, r_2,   r_I) V^{h\psi_3\psi_4}(  p_3,  p_4,-  p_I)+z^2
     V^{h\psi_1\psi_2}(   r_1, r_2,   r_I) V^{h\psi_3\psi_4}(  r_3,  r_4,-  r_I)\,.
     \end{aligned}
     \label{decomposeVV}
    \end{equation} 
The $t$- and $u$-channels are obtained by the exchanges $1\leftrightarrow4$, $1\leftrightarrow3$, respectively, and multiplying by a minus sign. Using the residue formula, one finds that  the linear terms in $z$ of \eqref{decomposeVV}  drop out and only the constant and quadratic terms in $z$  will contribute to the residue. Remarkably, when one evaluates  the quadratic term at the poles $z_\pm$, the denominator  $(z^+_{ij}-z^-_{ij})$ is cancelled, leaving in the numerator the product $z^+_{ij}z^-_{ij}$, which, thanks to the relation \eqref{zpluszminus}, will cancel the propagator. As a result, this  $z^2$ dependence in the expansion \eqref{decomposeVV} gives rise to a \textit{contact term}!

Recall that for the ALT shift, the shifted momentum is proportional to the (spin-1) polarization vector of the external particle, namely, for a gravitino of helicity $\psi_\mu^{++}=u^+ \varepsilon_\mu^+$ we have $r_\mu\propto \varepsilon_\mu^+$. This relation helps to simplify the expression and eliminate all $r$ dependence in the final result. Therefore, we arrive at the contact term
\begin{equation}
\begin{aligned}
 A_\text{contact}= \frac{c_h^2}{8}\kappa^2
    &\left[\bar\psi_1 ^\mu \gamma^\nu \psi _2^\alpha \bar{\psi}_{3\mu }\gamma_\nu \psi_{4\alpha} -\bar\psi_1 ^\mu \gamma^\nu \psi _2^\alpha \bar{\psi}_{3\alpha }\gamma_\nu \psi_{4\mu}+\bar\psi_1 ^\mu \gamma^\alpha \psi _2^\nu \bar{\psi}_{3\alpha }\gamma_\mu \psi_{4\nu}\right.\\
    &\left.+\bar\psi_1 ^\mu \gamma^\alpha \psi _2^\nu \bar{\psi}_{3\mu }\gamma_\nu \psi_{4\alpha} -\bar\psi_1 ^\mu \gamma^\nu \psi _2^\alpha \bar{\psi}_{3\nu }\gamma_\alpha \psi_{4\mu}-\bar\psi_1 ^\alpha \gamma^\nu \psi _2^\mu \bar{\psi}_{3\mu }\gamma_\alpha \psi_{4\nu}\right] \\
    &-(1\leftrightarrow3)-(1\leftrightarrow4)\,,
\end{aligned}    
\label{psicontactterm}
\end{equation}
which is antisymmetric under the exchange of any pair of fermions. Note that \eqref{psicontactterm} is written assuming transverse polarization vectors for all external legs. We can use the little-group covariance of the helicity amplitude to obtain the corresponding longitudinal amplitudes by applying spin raising or lowering operators which amounts to just replacing the $\psi_i$ in \eqref{psicontactterm} with longitudinal polarization vectors.  
Very importantly, when $c_h=\frac{1}{2}$, corresponding to the minimal coupling \eqref{mincoup32}, the contact term Eq.~\eqref{psicontactterm} is precisely equivalent to the four-fermion term in supergravity, up to terms that vanish on-shell \cite{Wess:1992cp}. 

In supergravity, the contact term \eqref{psicontactterm} can be obtained from the curved space covariant derivative of the gravitino, namely
\begin{equation}
    \nabla_\mu\psi_\nu=\left(\partial_\mu+\frac{1}{4}\omega_{\mu ab}\gamma^{ab}\right)\psi_\nu\,.
\end{equation}
The contorsion tensor $K_{\mu\nu\rho}$  enters the spin connection $\omega_{\mu ab}$ with
\begin{equation}
    \omega_{\mu a b }\equiv \omega_{\mu a b }(e)+K_{\mu a b } \,,
\end{equation}
where the torsionless part is 
\begin{equation}
\omega_\mu^{a b}(e)=2 e^{\nu[a} \partial_{[\mu} e_{\nu]}^{b]}-e^{\nu[a} e^{b] \sigma} e_{\mu c} \partial_\nu e_\sigma^c\,,
\end{equation}
and $K_{\mu a b }$ is bilinear in the gravitino field. Hence, in supergravity, from the kinetic term we obtain dimension six four-fermion interactions.

Amplitudes with longitudinal states can be derived from the all-transverse amplitude \eqref{psicontactterm} by replacing the polarization vectors. For instance,  with all longitudinal (+,+,+,+) helicities, the leading term is 
\begin{equation}
A_\text{contact}^{+,+,+,+}=    \frac{4c_h^2\kappa^2 E^6}{9m_{3/2}^4}(7-3 \cos2\theta)+\mathcal{O}(E^4)\,.
\label{contactgravexchange}
\end{equation}

The factorizable part of the amplitude with a kinematic pole arises from the constant in $z$ piece in Eq.~\eqref{decomposeVV}. 
Using \eqref{Vhpsipsiexp} and \eqref{simpTPIT}, the residue formula gives, for $s$-channel graviton exchange
\begin{equation}
    A_\text{s}=\frac{1}{s}T^{\mu\nu}_{3/2}(p_1,p_2)\Pi_{\mu\nu\alpha\beta} T_{3/2}^{\alpha\beta}(p_3,p_4)\,,
    \label{Asresult}
\end{equation}
and analogously for the $t$- and $u$-channels. It is obvious that the above result coincides with the amplitude obtained from using the Feynman rules. We can express \eqref{Asresult} in terms of polarization vectors and then derive all the longitudinal helicity amplitudes using spin raising/lowering operators. For  (+,+,+,+) helicity, the factorizable part of the amplitude with a kinematic pole is
\begin{equation}
    \begin{aligned}&
A_\text{s}^{+,+,+,+}=    \mathcal{O}(E^4)\,, 
\\&
A_\text{t}^{+,+,+,+}=      \frac{2c_h^2\kappa^2 E^6}{9m_{3/2}^4}(7+4 \text{ cos}\,\theta -3 \text{ cos}\,2\theta)+\mathcal{O}(E^4)\,,
\\&
A_\text{u}^{+,+,+,+}=     \frac{2c_h^2\kappa^2 E^6}{9m_{3/2}^4}(7-4\text{ cos}\,\theta-3 \text{ cos}\,2\theta)+   \mathcal{O}(E^4)\,.
    \end{aligned}\label{stugravexchange}
\end{equation}
Summing up \eqref{contactgravexchange} and  \eqref{stugravexchange} (with an additional minus sign in front of $A_\text{t} $ and $A_\text{u} $), the leading $\mathcal{O}(E^6)$ term cancels, which agrees with the result from minimal supergravity \cite{Antoniadis:2022jjy}. This cancellation is also similar to that observed in $WW\rightarrow WW$ scattering. 

The next order, corresponding to $\mathcal{O}(E^4)$, is non-vanishing
\begin{equation}
  A_\text{grav}^{+,+,+,+}=  A_\text{s}^{+,+,+,+}-A_\text{t}^{+,+,+,+}-A_\text{u}^{+,+,+,+}+A_\text{contact}^{+,+,+,+}= -\frac{16c_h^2\kappa^2 E^4}{3m_{3/2}^2}+\mathcal{O}(E^2)\,.
\label{gravpppp}
\end{equation}
This amplitude would lead to tree-level unitarity violation at a scale of order $\sqrt{m_{3/2} M_P}$ which can be much below $M_P$. To increase the scale of unitarity violation requires 
cancelling the $\mathcal{O}(E^4)$ term which can be achieved by adding new degrees of freedom to the theory. 
Here we consider introducing a scalar $S$ and a pseudoscalar $P$, and as we will see, their couplings are to be fixed by requiring tree-level unitarity, i.e.~the energy growth in the amplitude is no larger than $\mathcal{O}(E^2)$ so that the scale of unitarity violation is increased to be of order $M_P$.  
 
Let us first include a scalar particle. The scalar-gravitino-gravitino interaction in \eqref{psipsicouplings} contains no derivative, and thus the three-point amplitude is $z$-independent after the shift
\begin{equation}
       V^{S\psi_1\psi_2}(\hat{p}_1,\hat{p}_2,\hat{p}_I) V^{S\psi_3\psi_4}(\hat{p}_3,\hat{p}_4,-\hat{p}_I)=
     V^{S\psi_1\psi_2}(   p_1, p_2,   p_I) V^{S\psi_3\psi_4}(  p_3,  p_4,-  p_I)\equiv 
V^{S\psi_1\psi_2}  V^{S\psi_3\psi_4}, \end{equation}
so that the amplitude only has a factorizable contribution with a kinematic pole and no contact term. It is straightforward to derive the scalar contribution to the amplitude by using \eqref{fact-and-cont} to give
\begin{equation}
\begin{aligned}
A_\text{scalar}^{(\lambda_1 \lambda_2 \lambda_3\lambda_4)}=&\sum_{(i,j) = (1,2), (1,3), (1,4)}\frac{1}{p_{ij}^2-m_S^2} \left[
	\frac{z^+_{ij}-z^-_{ij}}{z^+_{ij} - z^-_{ij}}  V^{S\psi_i\psi_j}(   p_i, p_j,   p_{ij}) V^{S\psi_k\psi_l}(  p_k,  p_l,-  p_{ij})
	\right]\,,\\
 = &\sum_{(i,j) = (1,2), (1,3), (1,4)}\frac{1}{p_{ij}^2-m_S^2} 
	 V^{S\psi_i\psi_j} V^{S\psi_k\psi_l}\,, \\
  =&\sum_{(i,j) = (1,2), (1,3), (1,4)}\frac{c_S^2m_{3/2}^2\kappa^2}{p_{ij}^2-m_S^2}   \Bar{\psi}^\mu_i {\psi}_{j\mu}  \Bar{\psi}^\nu_k {\psi}_{l\nu}\,,
  \label{eq:scalaramp}
\end{aligned}
\end{equation}
where the sum is over the $s,t,u$ channels and again, there is an extra minus sign in the $t$- and $u$-channel contributions. 
The last line in \eqref{eq:scalaramp} depends only on polarization vectors and thus holds for any helicity. 

Next we evaluate the scalar contribution in the center-of-mass frame. For the (+,+,+,+) helicity amplitude we obtain
\begin{equation}
    A_\text{scalar}^{+,+,+,+}=\frac{16c_S^2\kappa^2E^4}{9m_{3/2}^2}+\mathcal{O}(E^2)\,.
\label{scalarpppp}
\end{equation}
The total order $\mathcal{O}(E^4)$ term arising from \eqref{gravpppp} and \eqref{scalarpppp} can be cancelled by appropriately choosing the coupling coefficients $c_S$ and $c_h$. However, including  only a real scalar $S$ does not suffice to ensure unitarity for all helicity amplitudes. For instance, the  (+,+,+,$-$)  amplitude is
\begin{equation}
 \begin{aligned}&
     A_\text{grav}^{+++-}=
     \frac{4c_h^2\kappa^2E^3 }{3m_{3/2}}\text{sin}\,2\theta +\mathcal{O}(E)\,,\\&
     A_\text{scalar}^{+++-}=-\frac{8c_S^2 \kappa^2E^3}{9m_{3/2}} \text{sin}\,2\theta+\mathcal{O}(E)\,.
 \end{aligned}
 \label{eq:pppmAgrav}
\end{equation}
To cancel the (+,+,+,$-$) amplitude in \eqref{eq:pppmAgrav} at $\mathcal{O}(E^3)$, requires imposing a separate condition on the coupling coefficients $c_S$ and $c_h$ which cannot be simultaneously satisfied with the condition needed for the cancellation at $\mathcal{O}(E^4)$ for the (+,+,+,+) amplitude. Therefore, a new degree of freedom is needed and we will introduce a pseudoscalar particle $P$.\,\footnote{Note that introducing another scalar field $S^\prime$ with the same interaction $\psib^\mu\psi_\mu S^\prime$  amounts to rescaling $c_S^2$ in the amplitudes, and therefore does not solve the problem.}

For pseudoscalar exchange, the vertex \eqref{psipsicouplings} has one derivative, but one can show that the $z^2$ dependence, as in \eqref{decomposeVV}, vanishes, i.e.~there is no contact term arising from the pseudoscalar interaction. This can be seen from
\begin{equation}
     V^{P\psi_1\psi_2}(   r_1, r_2,   r_I) =- c_P\kappa\,\varepsilon^{\mu\nu\alpha\beta } r_{I\mu}  \psib_{1\nu} \gamma_\alpha\psi_{2\beta}\,.
\end{equation}
Note that  $r_{I\mu}=r_{1\mu}+r_{2\mu}$ and $r_{1,2\mu}\propto \varepsilon_{1,2\mu}$  for all transverse external states.  Since $\psib_{ 1\mu}=\varepsilon_{ 1\mu} \bar{v}_1$, $\psi_{2\mu}=\varepsilon_{ 2\mu} u_2$, we can then conclude by antisymmetry of the Levi-Civita tensor that $ V^{P\psi_1\psi_2}(   r_1, r_2,   r_I) =0$. This conclusion carries over to the longitudinal states by using the spin raising or lowering operator. Alternatively, a simpler way to see this is by noticing that the coupling $ i 
\varepsilon^{\mu\nu\alpha\beta } \partial_\mu P \psib_\nu \gamma_\alpha\psi_\beta $ is equivalent on-shell to $P\psib_\nu \gamma_5  \psi^\nu$ which contains no derivative, and thus no contact term will arise.

The four-point amplitude from pseudoscalar exchange contains only the factorizable part with a kinematic pole
\begin{equation}
\begin{aligned}
    A_\text{p-scalar}^{(\lambda_1 \lambda_2 \lambda_3\lambda_4)}&=\sum_{(i,j) = (1,2), (1,3), (1,4)} \frac{1}{p_{ij}^2-m_P^2}V^{P\psi_1\psi_2}(   p_1, p_2,   p_I) V^{P\psi_3\psi_4}(   p_3, p_4, -  p_I)\,,
    \\&=\sum_{(i,j) = (1,2), (1,3), (1,4)}\frac{ c_P^2\kappa^2}{p_{ij}^2-m_P^2}
\varepsilon^{\mu\nu\alpha\beta } p_{I\mu}  \psib_{1\nu} \gamma_\alpha\psi_{2\beta}\varepsilon^{\rho\sigma\gamma\delta } (-p_{I\rho})  \psib_{3\sigma} \gamma_\gamma\psi_{4\delta}\,,
    \end{aligned}\label{eq:pscalaramp}
\end{equation}
where the sum is again over the $s,t$ and $u$ channels. For the (+,+,+,+) and (+,+,+,$-$) helicity amplitudes, the pseudoscalar contribution to the amplitude is
\begin{equation}
\begin{aligned}&A_\text{p-scalar}^{+,+,+,+}=\frac{64c_P^2\kappa^2E^4}{9m_{3/2}^2}+\mathcal{O}(E^2)\,,\\& A_\text{p-scalar}^{+++-}=0\,.
\end{aligned}
\end{equation}

Requiring the unitarity cutoff to be of order the Planck scale amounts to requiring that the leading high-energy behavior of the amplitude is of $\mathcal{O}(E^2)$. This requires cancelling all contributions at $\mathcal{O}(E^3)$ and $\mathcal{O}(E^4)$. Hence, for the (+,+,+,+)  helicity amplitude we obtain the constraint
\begin{equation}
    4c_P^2 +  c_S^2-3 c_h^2=0\,,
    \label{unitarity1}
\end{equation}
while for the (+,+,+,$-$) amplitude,  cancelling the $\mathcal{O}(E^3)$ contribution implies
\begin{equation}
     3c_h^2  -2c_S^2 =0\,.
\label{unitarity2}\end{equation}
As discussed earlier,  to satisfy \eqref{unitarity1} and \eqref{unitarity2}, both scalar \textit{and} pseudoscalar contributions are required, and in fact these two conditions suffice to guarantee  tree-level unitarity in all helicity configurations, so no new particles are necessary. Solving \eqref{unitarity1} and \eqref{unitarity2}, the coupling coefficients are related by $4  c_P^2=  c_S^2$, $3 c_h^2=8 c_P^2$. Ignoring the irrelevant relative signs, the on-shell three-point vertices with two gravitinos that 
guarantee a unitarity cutoff at the Planck scale are 
\begin{equation}\kappa
 c_P\left[2 m_{3/2}   \psib^\mu\psi_\mu S+ i 
\varepsilon^{\mu\nu\alpha\beta } \partial_\mu P \psib_\nu \gamma_\alpha\psi_\beta +2\sqrt{\frac{2}{3}} h_{\mu\nu }T^{\mu\nu}_{3/2}\right]\,.
\end{equation}
When $ c_P=\frac{1}{4}\sqrt{\frac{3}{2}}$, the above couplings are equivalent on-shell to the Polonyi model, and the four-fermion contact term \eqref{psicontactterm} also matches, as mentioned earlier. 

To summarize, let us emphasize that we arrived at the (on-shell) Polonyi model from the bottom up without recourse to Lagrangians or Feynman diagrams, assuming the following input information:
\begin{enumerate}[{(i)}]
    \item Particle content: (Majorana) gravitino, graviton, scalar, pseudoscalar. There is no spin-1/2 particle.
    \item The three-point interaction terms have dimension $<6$.
    \item The Ward identity for the four-point amplitude is satisfied in the massless limit.
    \item The tree-level unitarity bound for gravitino scattering is of order $M_P$.
\end{enumerate}
Assuming the particle content in (i), the general on-shell, three-point amplitudes were obtained from consistency with little-group and Lorentz-group transformations. The inputs (ii) and (iii) were then necessary in order to ensure constructibility under the ALT shift. 
Furthermore, the three-point interactions were chosen to have a smooth massless limit at high energy consistent with ``IR unification" or current conservation.

Finally, note that compared to the Feynman diagram approach which uses off-shell interactions, the advantage of the on-shell method is that all on-shell vanishing interactions are absent in the calculation, and all on-shell equivalent interactions are packaged into a single three-point amplitude. In particular, as discussed in \cite{Antoniadis:2022jjy}, one may attempt to cancel the $\mathcal{O}(E^4)$ term in gravitino scattering with a gauge boson exchange, but after expanding the supergravity Lagrangian it was shown that such couplings do not exist. This conclusion agrees with the on-shell calculation, since we explicitly showed in section \ref{s3232}, that there is no on-shell coupling of the (Majorana) gravitino-gravitino-gauge boson, and therefore the gauge boson does not contribute to gravitino scattering.

\subsection{The massless limit}

After constructing the massive scattering amplitudes, it is interesting to explicitly check that one recovers naturally the massless amplitude, as expected from the equivalence theorem~\cite{Casalbuoni:1988kv,Casalbuoni:1988qd}. From  Eq.~\eqref{masslesspsipsis1},\eqref{masslesspsipsis2} and \eqref{masslesspsipsih} we see that the high-energy limit of the (pseudo-) scalar interaction is dominated by a spin 1/2-spin 3/2-spin 0 coupling, whereas the graviton coupling at high energy reproduces the graviton coupling to a pair of spin 1/2 or spin 3/2 with opposite helicity. These massless three-point amplitudes can be readily employed to reproduce the scattering process 
$ \frac{3}{2},  -\frac{3}{2} \rightarrow \frac{1}{2},  -\frac{1}{2}$, and match to the massless limit of the gravitino scattering amplitudes computed in section~\ref{fourpointsec}.

First, assuming the helicity $\frac{3}{2}$, $-\frac{3}{2}$, $\frac{1}{2}$, $-\frac{1}{2}$ states are labelled as $1,2,3,4$, respectively, and the intermediate particle as $I$, the helicity amplitude $ \frac{3}{2},  -\frac{3}{2}, \frac{1}{2},  -\frac{1}{2}$ from Eqs.~\eqref{psicontactterm}, \eqref{stugravexchange}, \eqref{eq:scalaramp}, \eqref{eq:pscalaramp}  gives  
\begin{equation}
A_4^{++,--,+,-}=    \frac{1}{3}c_S^2\kappa^2E^2(\cos 2\theta -1)
=-\frac{2}{3}c_S^2\kappa^2E^2\sin^2\theta\,,
\label{masslesspmpm}
\end{equation}
where all masses are sent to zero and we have used the relations $4  c_P^2=  c_S^2$, $3 c_h^2=8 c_P^2$ for the couplings. In fact, only the scalar and pseudoscalar exchanges contribute to $\mathcal{O}(E^2)$. 

On the other hand, we may also take the massless limit of the massive three-point amplitudes, before on-shell constructing the four-point amplitude. The graviton mediation gives rise to the $s$-channel with
\begin{center}
\begin{tabular}{ccccc}
    \adjustbox{valign=m}{\includegraphics[width=0.2\textwidth]{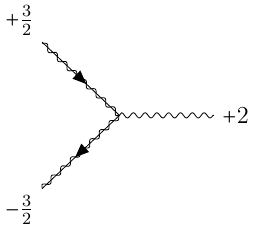}}  & $=-c_h\kappa\frac{[1I]^5}{[2I][12]^2}$&\qquad &
    \adjustbox{valign=m}{\includegraphics[width=0.2\textwidth]{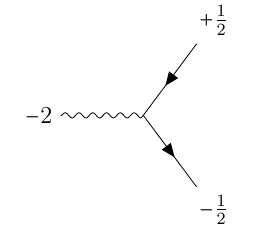}} &$ 
    =-c_h\kappa\frac{\left<4I\right>^3\left<3I\right>}{2\left<34\right>^2}$ \\
    \adjustbox{valign=m}{\includegraphics[width=0.2\textwidth]{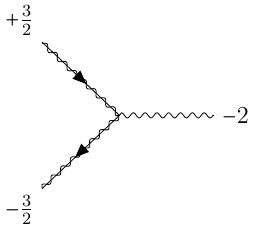}} &$=c_h\kappa\frac{\left<2I\right>^5}{\left<1I\right>\left<21\right>^2}$& \qquad&
    \adjustbox{valign=m}{\includegraphics[width=0.2\textwidth]{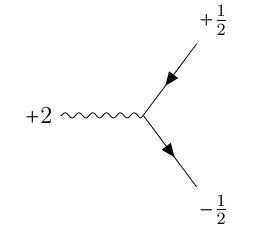}}  &$
 =c_h\kappa\frac{[3I]^3[4I]}{2[34]^2}$
\end{tabular}
\end{center}
The (pseudo-)scalar exchange corresponds to the $t$-channel:
\begin{center}
\begin{tabular}{cc}
    \adjustbox{valign=m}{\includegraphics[width=0.2\textwidth]{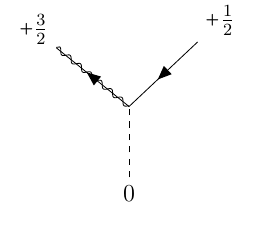}} &$=-c_0^{(+)}\frac{[13]^2[I1]}{[3I]}$\\ \adjustbox{valign=m}{\includegraphics[width=0.2\textwidth]{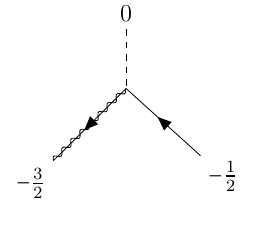}} &$ 
    =-c_0^{(-)}\frac{\left<24\right>^2\left<I2\right>}{\left<4 I\right>}$
\end{tabular}
\end{center}
where the coefficients are obtained from the massless limit of \eqref{psipsicouplings}
\begin{equation}
         c_0^{(+)}=\frac{\kappa}{\sqrt{3}}
         \left(-c_S-i2c_P \right)\,,
    \quad\quad  c_0^{(-)}=\frac{\kappa}{\sqrt{3}}\left(-c_S+i2c_P\right)\,.
\end{equation}

With the above information, we can construct the four-point amplitude using the recursion relation.  In the massless limit, the ALT shift amounts to shifting all momenta by their transverse polarization vectors. However, the shifted momentum can now be a generic massless reference vector, and not necessarily the polarization vector, because the on-shell condition is trivially satisfied in the massless case. One has more freedom in choosing the all-line shift as long as the total momentum is conserved. In addition, as we have briefly discussed for Compton scattering, the large-$z$ behavior is improved since the massless polarization vector scales as $1/z$. Each massless spin-3/2 state then brings an extra $1/z$ dependence, causing the boundary term to vanish, 
as long as we restrict to dimension $\leq5$ operators.\footnote{For a detailed discussion on massless all-line shifts, see for example \cite{Cohen:2010mi}.} 

For convenience, we choose the all-line shift
\begin{equation}
    \begin{aligned}
        & \left|\hat{1}\right>=\left|1\right>+z_1\left|4\right>,\quad \left|\hat{4}\right]=\left|4\right]+z_4\left|1\right]\,,
        \\&\left|\hat{3}\right>=\left|3\right>+z_3\left|2\right>,\quad \left|\hat{2}\right]=\left|2\right]+z_2\left|3\right]\,.
    \end{aligned}
    \label{masslessshift}
\end{equation}
The on-shell condition is straightforward, and there is always enough freedom in $z_{1,2,3,4}$ to satisfy momentum conservation. We now proceed to compute the product of  three-point amplitudes, which results in the following two contributions
\begin{itemize}
    \item 
$s$-channel:
\begin{equation}
    A_{3s}A_{3s}=\frac{1}{2}\kappa^2c_h^2 [13]\left<24\right>\left(\frac{[14]\left<24\right>^2}{\left<14\right>}+\frac{[13]^2\left<23\right>}{\left[23\right]}\right)\,.
\end{equation}

\item
$t$-channel:
\begin{equation}
  A_{3t}A_{3t}=-c_0^{(+)}c_0^{(-)} \frac{[13]^2\left<24\right>^2\left<23\right>}{\left<14\right>}\,.
\end{equation}
\end{itemize}
We have included a minus sign in the $t$-channel accounting for fermion exchange. Using the shift \eqref{masslessshift} we have simply $ \hat{A}_3\hat{A}_3= A_3A_3$, so the residue formula along with the identity \eqref{zpluszminus} yields
\begin{equation}
    A_4=-\frac{A_{3s}A_{3s}}{s}-\frac{A_{3t}A_{3t}}{t}\,.
\end{equation}
Evaluating the above expression in the center-of-mass frame, we recover 
\begin{equation}
\begin{aligned}
    A_4&=\frac{1}{2}\kappa^2 c_h^2E^2(3+4\cos\theta+\cos 2\theta)-\frac{4}{3}\kappa^2 c_S^2 E^2(1+\cos\theta)\\&  = -\frac{2}{3}c_S^2\kappa^2E^2\sin^2 \theta\,,
\end{aligned}
\label{eq:gluemassless3pt}
\end{equation}
which is exactly Eq.~\eqref{masslesspmpm}. 

To summarize,  we have  computed  the helicity amplitude $ \frac{3}{2},  -\frac{3}{2}\rightarrow \frac{1}{2},  -\frac{1}{2}$ in two different ways: first, we constructed the massive gravitino scattering amplitude using the ALT shift as discussed in section~\ref{fourpointsec}, and then we took the massless limit of the amplitude.
 On the other hand, using the massless limit of the massive three-point amplitudes,  in Eq.~\eqref{masslesspsipsis1}-\eqref{masslesspsipsis2}, and \eqref{masslesspsipsih} with appropriate coupling constants, we on-shell construct the massless four-point amplitude using an all-line shift.  
The two results in \eqref{masslesspmpm} and \eqref{eq:gluemassless3pt} match, as expected. 

As a final comment, we expect the same matching to hold for all longitudinal amplitudes, including the $\frac{1}{2},\frac{1}{2},\frac{1}{2},\frac{1}{2}$ amplitude. The on-shell construction of massive amplitudes in section~\ref{fourpointsec} gives the $\mathcal{O}(E^2)$ amplitude when all masses are set to zero. By the equivalence theorem, this should correspond to the massless spin-1/2 scattering amplitude. This is indeed the case, when Feynman rules are used with the minimal coupling $\frac{\kappa}{2} T^{\mu\nu}_{1/2}h_{\mu\nu}$, where $T^{\mu\nu}_{1/2}$ is the energy-momentum tensor for massless spin 1/2, and the four-(chiral-)fermion contact term from supergravity is added to the massless theory. 
However, the massless four-point amplitude cannot be constructed on-shell using the all-line shift. Since the interactions have dimension 5, the large-$z$ behavior is $z^0$, and the Ward identity is no longer useful, because the spin-1/2 external states have no associated polarization vector.\footnote{Recall that we can impose the Ward identity on spin-1 \cite{Ema:2024rss} and spin-3/2 amplitudes to eliminate $z^0$ terms, because the polarization vector $\varepsilon_\mu$ that can be associated with external states is proportional to the massless shifted momentum.} The non-constructibility of this class of massless amplitudes has also been argued in Ref.~\cite{McGady:2013sga} by pole counting.

\section{Conclusion}  
\label{sec:concl}

In this paper, we have used the massive spinor-helicity formalism, introduced in Ref.~\cite{Arkani-Hamed:2017jhn}, to construct all three-point interactions of massive (Majorana) spin-3/2 particles that contain a graviton and particles with spin $\leq 1$. These amplitudes are only constrained by little-group and Lorentz group transformations. The interactions involving two spin-3/2 particles with a gauge boson vanish, as expected for a Majorana particle, while those coupled to the graviton reproduce the usual gravitational minimal coupling obtained from the energy-momentum tensor. The couplings involving just one spin-3/2 particle may be phenomenologically useful in the study of spin-3/2 decays, since no Lagrangian has been used and the lifetime only depends on the number of parity even and odd interactions. In the massless limit, the three-point interactions containing the helicity $\pm \frac{1}{2}$ components of the spin-3/2 particle reduce to the massless amplitudes containing a chiral fermion, by requiring there are no mass poles at UV scales. This reveals an ``IR unification" manifesting as the superHiggs effect as discussed in \cite{Arkani-Hamed:2017jhn} or alternatively requiring current conservation~\cite{Cucchieri:1994tx}.

The three-point interactions are then used to on-shell construct massive four-point amplitudes. In particular, we consider the four-point massive spin-3/2 amplitude and use the ALT shift introduced in Refs.~\cite{Ema:2024rss,Ema:2024vww} to combine the graviton three-point amplitude assuming the Ward identity. An interesting consequence of this momentum shift is that the four-point gravitino contact interaction, that appears in the $N=1$ supergravity Lagrangian, is reproduced. The leading high-energy behavior of the massive spin-3/2 four-point amplitude scales as ${\cal O}(E^6)$, which becomes ${\cal O}(E^4)$ thanks to the contact term arising from the recursion relation. This behavior is further  ameliorated to ${\cal O}(E^2)$ by including the scalar and pseudoscalar exchange diagrams in the on-shell recursion, reproducing the well-known perturbative unitarity violation at the Planck scale, solely from on-shell methods. As a final application of on-shell methods, we check the massless limit of the four-point massive spin-3/2 amplitude for the $\pm \frac{3}{2}, \pm\frac{1}{2}$ helicity case. The massless version of this amplitude can now be independently computed using an all-line-shift, which has good large-$z$ behavior, and matches the result obtained from the massless limit of the massive amplitude.

We emphasize that our results are obtained without the use of a Lagrangian or Feynman diagrams and provide a new perspective on massive (Majorana) spin-3/2 interactions. The results can be straightforwardly generalized to the case of a Dirac spin-3/2 particle with an electromagnetic coupling. It would be interesting to extend the analysis to dimension 6 operators and study under which conditions the amplitude can be constructible, or perform a more systematic study of perturbative unitarity constraints with different external states. Finally, our results could be useful to further understand the double copy of amplitudes in massive supergravity.

\section*{Acknowledgements}

We thank Yohei Ema and Anthony Guillen for useful discussions. This work is supported in part by the Department of Energy under Grant No. DE-SC0011842 at the University of Minnesota.

\appendix

\section{Conventions and useful relations}

\subsection{Spinor helicity formalism}
\label{app:conv}

We present relevant details of the spinor helicity formalism, including our conventions and identities that are useful to obtain our results.
We assume the Minkowski metric $\eta_{\mu\nu}={\rm diag}(+1,-1,-1,-1)$ and spinor indices are raised and lowered by $\varepsilon_{\alpha\beta}$, $\varepsilon_{\dotalpha\dotbeta}\equiv (\varepsilon_{\alpha\beta})^*$ where
\begin{equation}
    \varepsilon_{\alpha\beta}=-\varepsilon^{\alpha\beta}=\left(\begin{array}{cc}
         0&-1  \\
         1&0 
    \end{array}\right)\,.
    \label{defepsilon}
\end{equation}
The gamma matrices are defined to be
\begin{equation}
    \gamma^\mu=\left(\begin{array}{cc}
        0 &\sigma^\mu _{\alpha\dotalpha}  \\
         \bar{\sigma}^{\mu\beta\dotbeta}&0 
    \end{array}\right),\quad   \gamma_5=\left(\begin{array}{cc}
        -I_2 &0  \\
        0& I_2 
    \end{array}\right)\,,
\end{equation}
where $\sigma^\mu=(I_2,\sigma^i)$ and $\bar{\sigma}^\mu=(I_2,-\sigma^i)$ with the Pauli matrices $\sigma^i$ and $I_2$ is a $2\times 2$ identity matrix.

In the massless case, with four-momentum $p_\mu$ satisfying $p^2=0$, implies that the momentum bispinor $p_{\alpha\dotalpha}\equiv p_\mu \sigma^\mu_{\alpha\dotalpha}$ has rank 1, and hence can be decomposed into chiral ($\lambda_\alpha$) and anti-chiral ($\Tilde{\lambda}_{\dotalpha}$) spinors
\begin{equation}
    p_{\alpha\dotalpha} 
    =\lambda_\alpha\Tilde{\lambda}_{\dotalpha}\equiv \left|p\right>_\alpha\left[p\right|_{\dotalpha},\qquad     \bar{p}^{\dotalpha\alpha}\equiv p_\mu \bar{\sigma}^{\mu\dotalpha\alpha} =\Tilde{\lambda}^{\dotalpha}\lambda^\alpha\equiv \left|p\right]^{\dotalpha}\left<p\right|^{\alpha}\,,
\end{equation} 
where $\lambda_\alpha$ (angle bracket) has helicity $-1/2$ and  $\tilde{\lambda}_{\dotalpha}$ (square bracket) has helicity $1/2$.  
The spinors with dotted and undotted indices transform respectively under the $(0,1/2)$, and $(1/2,0)$ representations of the $SL(2,\mathbb{C})$ group. In the massless case, the $U(1)$ little-group transformations are
\begin{equation}
    \lambda_\alpha \rightarrow e^{-i\xi }  \lambda_\alpha,\quad 
    \tilde{\lambda}_{\dotalpha} \rightarrow e^{i\xi }  \tilde{\lambda}_{\dotalpha}\,.
\end{equation}
It is obvious that $p_{\alpha\dotalpha}$ is invariant under this transformation. This allows a massless three-point amplitude to be uniquely determined by the helicities $h$ of the particles. When $h_1+h_2+h_3>0$ it is given by
\begin{equation}
    M^{h_1 h_2 h_3}=c_+ [12]^{h_1+h_2-h_3}[23]^{h_2+h_3-h_1}[31]^{h_3+h_1-h_2}\,,
    \label{masslesscplus}
\end{equation}
while for $h_1+h_2+h_3<0$ the amplitude is
\begin{equation}
     M^{h_1 h_2 h_3}=c_-\left<12\right>^{h_3-h_1-h_2}\left<23\right>^{h_1-h_2-h_3}\left<31\right>^{h_2-h_3-h_1}\,,
\label{masslesscminus}
\end{equation}
where $c_\pm$ are coupling coefficients. The amplitude expressed in terms of these spinor-helicity variables is manifestly little-group covariant. 

For massive particles ($p_\mu p^\mu=m^2$) the momentum $p_{\alpha\dotalpha}$ 
has rank 2 because $\text{det}(p)=m^2$.  
To describe these momenta, one introduces additional $SU(2)$ indices $I=1,2$, which are raised and lowered by $\varepsilon_{IJ}$ defined in \eqref{defepsilon}, The momentum can then be decomposed as~\cite{Arkani-Hamed:2017jhn}
\begin{equation}   
p_{\alpha\dotalpha}=\epsilon_{IJ}\lambda^I_\alpha\Tilde{\lambda}^J_{ \dotalpha}=\lambda^I_\alpha\Tilde{\lambda}_{I \dotalpha}\equiv \left|\bp\right>^I_\alpha\left[\bp\right|_{I\dotalpha},\qquad     \bar{p}^{\dotalpha\alpha} =\epsilon^{IJ}\Tilde{\lambda}_I^{\dotalpha}\lambda_J^{\alpha}=\Tilde{\lambda}_I^{\dotalpha}\lambda^{I\alpha}\equiv \left|\bp\right]_I^{\dotalpha}\left<\bp\right|^{I \alpha}\,.
\label{mom-decom}
\end{equation}
The spinors obey the Dirac equation
\begin{equation}
    p|\bp]=m\left| \bp\right>,\quad\bar{p}\left|\bp\right>=m\left| \bp\right],\quad \left[\bp\right|\Bar{p}=-m\left<\bp\right|,\quad \left<\bp\right|p=-m[\bp|\,,
    \label{diraceq-braket}
\end{equation}
which leads to the    identities
\begin{equation}
    \begin{aligned}
        &\left<\bp\right|^{I\alpha}\left|\bp\right>_{I\beta}  =m \delta^\alpha_\beta,\qquad \left[\bp\right|_{\dotbeta}^I\left|\bp\right]_I^{\dotalpha}  =-m\delta^{\dotalpha}_{\dotbeta}\,,\\& \left< \bp \bp\right>^{IJ}=-m\epsilon^{IJ},\qquad \left[ \bp \bp\right]^{IJ}=m\epsilon^{IJ} \,.
    \end{aligned}
\end{equation}
Note also the following useful Schouten identities
\begin{equation}
\begin{aligned}&
|\bthree][\bone\btwo] +|\bone][\btwo\bthree]+|\btwo][\bthree\bone]=0\,,\\&
\left|\bthree\right>\left<\bone\btwo\right> +\left|\bone\right>\left<\btwo\bthree\right>+\left|\btwo\right>\left<\bthree\bone\right>=0\,.
\end{aligned}
\label{schouten}
\end{equation}
Parity acts on $\lambda,\Tilde{\lambda}$ as 
\begin{equation}
    \lambda^I_\alpha \leftrightarrow \Tilde{\lambda}_I^{\dotalpha}\,,\qquad \Tilde{\lambda}^{I\dotalpha}\leftrightarrow -\lambda_{I\alpha}\,,
\end{equation}
which sends $ p_{\alpha\dotalpha}$ to $\bar{ p}^{\dotalpha\alpha}$.

A  particle of spin $s$ is constructed out of a combination of $2s$ chiral ($\lambda$) or anti-chiral ($\tilde{\lambda}$) spinors.  Starting with the simplest case, Dirac spinors are given in terms of $\lambda$, $\Tilde{\lambda}$  by
\begin{equation}
  \begin{aligned}
& u^I(p)=\left(\begin{array}{c}
           \lambda^I_\alpha\\
           \Tilde{\lambda}^{I\dotalpha}
    \end{array}\right)=\left(\begin{array}{c}
           \left|\bp\right>\\
           \left|\bp\right]
    \end{array}\right),\quad 
    v^I(p)=\left(\begin{array}{c}
           \lambda^I_\alpha\\
     -      \Tilde{\lambda}^{I\dotalpha}
    \end{array}\right)=\left(\begin{array}{c}\left|\bp\right>\\
    -       \left|\bp\right]
    
    \end{array}\right)\,,\\
    &\bar{u}_I(p)=\left(-\lambda_I^{\alpha}\quad \Tilde{\lambda}_{I\dotalpha}\right)=\left(-\left<\bp\right|\quad \left[\bp\right|\right),\quad \bar{v}_I(p)=\left(\lambda_I^{\alpha}\quad \Tilde{\lambda}_{I\dotalpha}\right)=\left(\left<\bp\right|\quad \left[\bp\right|\right)\,.
  \end{aligned} \label{uvlam}
\end{equation}
Using the Dirac equations \eqref{diraceq-braket}, we recover
\begin{equation}
\begin{aligned}
& (\slashed{p}-m) u^I(p)=0\,, \qquad(\slashed{p}+m) v^I(p)=0\,, \\
& \bar{u}_I(p)(\slashed{p}-m)=0\,, \qquad \bar{v}_I(p)(\slashed{p}+m)=0\,. \\
&
\end{aligned}
\end{equation}
The spin sum amounts to summing over $I$ indices, which results in the familiar relations: \begin{equation}
    \begin{aligned}
        \sum_I u^I \Bar{u}_I=\slashed{p}+m,\quad 
        \sum_I v^I \Bar{v}_I=\slashed{p}-m\,.
    \end{aligned}
\end{equation}
For a massive spin-1 particle of momentum $p$, the polarization vector is given by\footnote{The symmetrization of the $IJ$ indices is understood.}
\begin{equation}
\varepsilon_\mu^{IJ}=\frac{1}{\sqrt{2}m}
\left<\bp\right|\sigma_\mu\left|\bp\right],\qquad   \varepsilon^{IJ}_{\alpha\Dot{\alpha}}=
\frac{\sqrt{2}}{m}
\left| \bp\right>_\alpha\left[\bp\right|_{\Dot{\alpha}}\,,
\label{eq:massivepolvect}
\end{equation}
which can be decomposed into three polarizations $(+,-,0)$ for different $I$, $J$
\begin{equation}
\begin{aligned}    \varepsilon^+_{\alpha\dotalpha}\equiv\varepsilon_{\alpha\dotalpha}^{11},\quad 
    \varepsilon^-_{\alpha\dotalpha}\equiv\varepsilon_{\alpha\dotalpha}^{22},\quad 
    \varepsilon^0_{\alpha\dotalpha}\equiv\frac{1}{2}(\varepsilon_{\alpha\dotalpha}^{12}+\varepsilon_{\alpha\dotalpha}^{21})\,.
\end{aligned}    
\end{equation}
In the massless limit, they reduce to
\begin{equation}
    \varepsilon^+_{\alpha\dotalpha}= \frac{\sqrt{2}}{m}\left|\eta\right>_\alpha[\Tilde{\lambda}|_{\dotalpha}
    \rightarrow \sqrt{2}\frac{\left|\zeta\right>_\alpha[p|_{\dotalpha}}{\left<p\zeta\right>},\quad \varepsilon^-_{\alpha\dotalpha}= \frac{\sqrt{2}}{m}\left|\lambda\right>_\alpha[\Tilde{\eta}|_{\dotalpha}\rightarrow \sqrt{2}\frac{\left|p\right>_\alpha[\zeta|_{\dotalpha}}{\left[p\zeta\right]},\quad  
\varepsilon^0_{\alpha\dotalpha}\rightarrow 0\,,
\label{eq:masslesspolvect}
\end{equation}
where we have identified $\left|\eta\right>_\alpha/m\rightarrow \left|\zeta\right>_\alpha/\left<p\zeta \right>$, $\left[\eta\right|_{\dotalpha}/m\rightarrow \left[\zeta\right|_{\dotalpha}/\left[p\zeta \right]$, $\left|\lambda\right>_\alpha\rightarrow\left|p\right>_\alpha$, $[\Tilde{\lambda}|_{\dotalpha}\rightarrow\left[p\right|_{\dotalpha}$ and introduced $\zeta$ as a reference momentum. 

From these elements, one obtains the most general three-point amplitudes containing massive states, which are derived in section~\ref{sec:trilinear}. 
 In particular, for the one massless, two equal mass amplitude, one introduces the $x$-factor, defined in \eqref{xfactordef}. 
Denoting the two massive states by $\btwo$, $\bthree$, and the massless state by $1$, we obtain the following relations
\begin{equation}
  \begin{aligned}
  & x\left<\btwo\bthree\right>=\left<\btwo \varepsilon_1^+\bthree\right]+\left<\bthree \varepsilon_1^+\btwo\right],\quad \frac{1}{x}\left<\btwo\bthree\right>=\left<\btwo \varepsilon_1^-\bthree\right]+\left<\bthree \varepsilon_1^-\btwo\right]+\frac{1}{m}\left<1\btwo\right>\left<1\bthree\right>\,,
  \\& x\left[\btwo\bthree\right]=\left<\btwo \varepsilon_1^+\bthree\right]+\left<\bthree \varepsilon_1^+\btwo\right]+\frac{1}{m}[1\btwo][1\bthree]
 ,\quad \frac{1}{x}\left[\btwo\bthree\right]=\left<\btwo \varepsilon_1^-\bthree\right]+\left<\bthree \varepsilon_1^-\btwo\right]\,, 
 \\&x\left<1\btwo\right>=-\left[1\btwo\right],\quad \frac{1}{x} \left[1\btwo\right]=-\left<1\btwo\right>\,.
  \end{aligned}
  \label{xrelation}  
\end{equation}
They are derived by momentum conservation, Schouten identities and equations of motion.  Furthermore, they  lead to the identities
\begin{equation}
    \begin{aligned}
        &  x\left<\btwo\bthree\right>\times \left<1\btwo \right>= \left<\btwo\bthree\right>\times x\left<1\btwo \right>  \quad\Rightarrow\quad\left(\left<\btwo \varepsilon_1^+\bthree\right]+\left<\bthree \varepsilon_1^+\btwo\right]\right)\left<1\btwo\right>=-[1\btwo] \left<\btwo\bthree\right>\,,\\& \frac{1}{x}\left[\btwo\bthree\right]\times\left[1\btwo \right]= \left[\btwo\bthree\right]\times \frac{1}{x}\left[1\btwo \right]\quad\Rightarrow\quad\left(\left<\btwo \varepsilon_1^-\bthree\right]+\left<\bthree \varepsilon_1^-\btwo\right]\right)\left[1\btwo\right]=-\left<1\btwo\right>\left[\btwo\bthree\right]\,,
    \end{aligned}
    \label{polarelation}
\end{equation}
which are used in deriving the results in section~\ref{sec:trilinear}.

\subsection{Explicit kinematics}
\label{app:kin}   

For convenience, we provide the explicit kinematical formulae used in the computation of the amplitudes.
The four-momentum $p^\mu\equiv (E,\Vec{p})$, where the spatial component $\Vec{p}$ with magnitude $p\equiv |\vec{p}|$, is conveniently decomposed in spherical polar coordinates (with $0\leq \phi \leq 2\pi$ and $0\leq \theta \leq \pi$) as
\begin{equation}
    \Vec{p}=\left(p \sin \theta\cos\phi,p\sin\theta\sin\phi,p\cos\theta\right)\,.
\end{equation}
In the spinor basis, the momentum is expressed as $p_{\alpha\dotalpha}=\lambda_\alpha\Tilde{\lambda}_{\dotalpha}-\eta_\alpha\Tilde{\eta}_{\dotalpha}$, where the chiral and anti-chiral spinors are explicitly given in terms of the kinematic variables by
\begin{equation}
    \begin{aligned}
        &\lambda_\alpha=\sqrt{E+p}\left(\begin{array}{c}
             -s^* \\c
        \end{array}\right)\,,\qquad \Tilde{\lambda}_{\dotalpha}=\sqrt{E+p}\left(\begin{array}{c}
             -s \\c^*
        \end{array}\right)\,,\\&\eta_\alpha=\sqrt{E-p}\left(\begin{array}{c}
             c^* \\s
        \end{array}\right)\,,\qquad \Tilde{\eta}_{\dotalpha}=-\sqrt{E-p}\left(\begin{array}{c}
             c\\s^*
        \end{array}\right)\,,
    \end{aligned}\label{explicitkin}
\end{equation}
with the definitions
\begin{equation}
    s\equiv \sin \frac{\theta}{2}\,e^{\frac{i}{2}\phi },\quad c\equiv\cos \frac{\theta}{2}\,e^{\frac{i}{2}\phi}\,.
\end{equation}
The momentum bispinor then becomes
\begin{equation}
p_{\alpha \dot{\alpha}}=\left(\begin{array}{cc}
E-p \cos \theta & -p \sin \theta e^{-i \phi} \\
-p \sin \theta e^{i \phi} & E+p \cos \theta
\end{array}\right)\,.
\end{equation}
We choose the polarization vectors to be
\begin{equation}
    \begin{aligned}
        &\varepsilon^{\mu+}=-\frac{1}{\sqrt{2}}(0,\cos \theta\cos \phi-i \sin \phi,\cos \theta\sin \phi+i \cos \phi,-\sin \theta)\,,\\  &\varepsilon^{\mu-}=\frac{1}{\sqrt{2}}(0,\cos \theta\cos \phi+i \sin \phi,\cos \theta\sin \phi-i \cos \phi,-\sin \theta)\,,\\&\varepsilon^{\mu 0}=\frac{1}{m}(p,E\sin \theta \cos \phi,E\sin \theta \sin \phi, E \cos \theta)\,,
    \end{aligned}
\end{equation}
which satisfy the equations $p_\mu \varepsilon^{\mu,\pm,0}=0$, $\varepsilon^{\mu+}\varepsilon_\mu^{-}=1$, $\varepsilon^{\mu 0}\varepsilon_\mu^{0}=-1$, $\varepsilon^{\mu \pm}\varepsilon_\mu^{ \pm}=0$.
Note that we define all momenta to be incoming, in particular, our convention is that the final states have negative energy, so the  corresponding wavefunctions and  polarization vectors can be obtained by analytical continuation
\begin{equation}
    p^0\rightarrow -p^0 ,\quad p\rightarrow -p ,\quad \theta \rightarrow \theta,\quad \phi\rightarrow \phi\,,
\end{equation}or alternatively
\begin{equation}
      p^0\rightarrow -p^0 ,\quad p\rightarrow p ,\quad \theta \rightarrow \pi-\theta,\quad \phi\rightarrow \pi+\phi\,.
\end{equation}
In the spinor notation, we require\begin{equation}
    \lambda^I(-p)=-\lambda^I(p),\quad \tilde{\lambda}^I(-p)=\tilde{\lambda}^I(p)\,. 
\end{equation}

We evaluate the helicity amplitudes in the center-of-mass frame, where the angles are chosen to be
\begin{equation}\begin{aligned}
    &\theta_1=\pi,\quad \phi_1=0, \quad \theta_2=0,\quad \phi_2=0, \\&\theta_3=\theta-\pi,\quad \phi_3=0, \quad \theta_4=\theta,\quad \phi_4=0\,.
    \end{aligned}
    \label{comangles}
\end{equation}
 
\section{Constraints from current conservation}
\label{app:gaugeinv}  

In this appendix, we recover the interactions \eqref{psipsicouplings}, that were derived in section~\ref{gaugeinvsec} from ``IR unification", by instead imposing a constraint from current conservation. 

\subsection{General arguments}

In the study of higher-spin Compton amplitudes, Ref.~\cite{Cucchieri:1994tx} noticed that a necessary condition to have a unitary theory is
\begin{equation}
    \partial^\mu J_\mu =\mathcal{O}(m)\,,
    \label{eq:gaugeinvconstraint}
\end{equation}
where the current $J$ associated with a particle $\phi$ of spin $\geq1$ and mass $m$, is defined by
\begin{equation}
    J(\phi) \equiv \frac{\delta V}{\delta \phi}\,,
    \label{eq:currentdefn}
\end{equation}  
and $V$ represents a three-point vertex with off-shell $\phi$, and the two other particles in the vertex are on-shell.  
The constraint \eqref{eq:gaugeinvconstraint} forces the longitudinal component of the current to vanish when $m \rightarrow 0$. In other words, one recovers a gauge invariant theory. The Compton amplitude is schematically $J^\mu\Pi_{\mu\nu} J^\nu$, where 
$\Pi_{\mu\nu}$ denotes the massive propagator, $\Pi_{\mu\nu}\equiv \sum_\text{pol}\varepsilon_\mu\varepsilon_\nu /(p^2-m^2)$. The polarization sum in the numerator features momentum-dependent terms of the form $p_\mu\gamma_\nu/m $ and $p_\mu p_\nu/m ^2$ (associated with the longitudinal components), which would potentially violate unitarity at  high energy. However, if $ \partial^\mu J_\mu=\mathcal{O}(m)$, such terms from the polarization sum will always have a contraction with the current, leading to an amplitude which depends on the mass parameter $m$. More generally, the divergence of the current can depend on other masses of the theory, not necessarily the higher spin mass, as long as our goal is to cancel the momenta in the polarization sum.  

We can impose this constraint on the example of dimension $\leq 5$ operators with two spin-3/2 particles.   It can be verified that the minimal coupling vertex \eqref{gravitinotmunu} with off-shell spin 3/2 satisfies the constraint \eqref{eq:gaugeinvconstraint}. The current is given by
\begin{equation}
  \begin{aligned}
  J^\mu=&\frac{1}{8}\kappa \partial_\lambda h_{\rho\sigma}\varepsilon^{\mu\nu\rho\tau}\gamma_5\gamma_\sigma\gamma^\tau \gamma^\lambda\psi_\nu -\frac{1}{   4}\kappa h_{\rho\tau}\varepsilon^{\mu\nu\rho\lambda}\gamma_5\gamma^\tau\partial_\lambda \psi_\nu   +\mathcal{O}(m_{3/2})\,,
  \end{aligned}  
\end{equation}
and therefore, upon applying the equations of motion we obtain
\begin{equation}
 \partial^\mu   J_\mu=\mathcal{O}(m_{3/2})\,,
\end{equation}
where  $\bar{\psi}_\mu$ is taken to be off-shell, while its conjugate $\psi_\mu$ and the graviton $h_{\mu\nu}$ are on-shell. The corresponding three-point amplitude obtained from the interaction \eqref{gravitinotmunu} is then \eqref{mincouponshell}. 

One may wonder whether there exist other couplings satisfying gauge invariance in the massless limit, which lead to a different three-point amplitude compared to \eqref{mincouponshell}. 
In appendix~\ref{subc1}, we will show that the minimal coupling is indeed the only solution as long as the dimension of the operator is no larger than 5, i.e.~the condition of gauge invariance imposes the gravitational   coupling to be the minimal one.

As for vertices containing the scalar, the only dimension $\leq5$ on-shell interaction term is $\psib^\mu\psi_\mu S$, yielding a current $J_\mu \propto \psi_\mu S$ whose divergence does not vanish in the massless limit. In this case, one can explicitly introduce a mass-dependent coupling: 
$m_{3/2}  \kappa\, \psib^\mu\psi_\mu S$. 
In the massless limit, this operator will then simply vanish. 
The pseudoscalar interaction term of interest is $\varepsilon^{\mu\nu\alpha\beta } \partial_\mu P \psib_\nu \gamma_\alpha\psi_\beta$, which has a coupling of mass dimension $-1$, that we take to be proportional to $\kappa$. The corresponding current is then  
$J_\mu\propto \varepsilon_{\nu\mu\alpha\beta } \partial^\nu P  \gamma^\alpha\psi^\beta$, and therefore $\partial^\mu J_\mu\propto\mathcal{O}(m_{3/2})$ using the equation of motion.  
Thus, we find again the three-point interactions given in \eqref{psipsicouplings} involving two spin-3/2 particles. 

Similar arguments apply to couplings that contain one spin-3/2 particle.  For all interactions presented in Eq.~\eqref{dim5scalarfermion}, \eqref{eq:evenpdim5}-\eqref{odddim5}, one can always add a term that vanishes on-shell, so that the corresponding current is conserved in the massless limit. First, the scalar interaction in \eqref{dim5scalarfermion} corresponds to the current with divergence
\begin{equation}
   \begin{aligned}
\partial^\rho
\frac{\delta}{\delta \psib^\rho} (\psib^\mu \gamma^{\nu\alpha}\chi\varepsilon_{\mu\nu\alpha\beta}\partial^\beta P)
&=\partial^\mu (\gamma^{\nu\alpha}\chi\varepsilon_{\mu\nu\alpha\beta}\partial^\beta P)\,,\\
&=\gamma^{\nu\alpha}(\partial^\mu \chi)\varepsilon_{\mu\nu\alpha\beta}\partial^\beta P\,,\\
&=2m_\chi \gamma^5 \gamma_\beta \chi \partial^\beta P -2i\gamma^5 \partial^\mu \chi\partial_\mu P\,.
   \end{aligned}
   \label{C5eq} 
\end{equation}
The last term does not vanish in the massless limit, but it can be compensated by adding $- i\partial^\mu \psib_\mu \gamma^5 \chi P +\text{h.c.}$ (note that $\partial^\mu \psib_\mu=0$ on-shell), which gives
\begin{equation}
    \partial^\rho\frac{\delta}{\delta\psib^\rho}
    (-i\partial^\mu \psib_\mu \gamma^5 \chi P)
    =2i \gamma^5 \partial^\mu \chi\partial_\mu P+i (m_\chi^2+m_P^2)\gamma^5 \chi P\,.
    \label{eq:onshelldpsiP}
\end{equation}
Combining \eqref{C5eq} and \eqref{eq:onshelldpsiP}, the first term in \eqref{eq:onshelldpsiP} can now cancel the last term in \eqref{C5eq}, giving rise to a divergence that vanishes when $m_\chi,m_P\rightarrow0$. Similarly, for the scalar coupling $i\psib^\mu \partial_\mu \chi S +\text{h.c.}$, one can add $\frac{1}{2}i \partial^\mu\psib_\mu \chi S+\text{h.c.}$, so that the current is conserved in the massless limit. 

For the spin 1/2-spin 3/2-gauge boson coupling, we find that the term $i\psib^\mu \gamma_{\alpha\beta }\gamma_\mu \chi F^{\alpha\beta}+\text{h.c.}$, is equivalent on-shell to $i\psib^\mu \gamma^\nu \chi  F_{\mu\nu}+\text{h.c.}$, and satisfies
\begin{equation}
\partial^\mu J_\mu=\partial^\rho
\frac{\delta}{\delta\psib^\rho}
(i\psib ^\mu \gamma_{\alpha\beta}\gamma_\mu \chi F^{\alpha\beta} )
= m_\chi \gamma_{\alpha\beta} \chi F^{\alpha\beta},
\end{equation}
which is massless in the $m_\chi\rightarrow0$ limit. Note that the above interaction coincides with the gravitino-gaugino-gauge boson vertex the supergravity Lagrangian.
Therefore, we recover the on-shell interactions in 
\eqref{eq:psicouplings}.

\subsection{Proof for the graviton coupling}
\label{subc1}

We focus on interactions of dimension $\leq5$ involving two massive spin-3/2 states and one graviton.  We will show that maintaining gauge invariance in the massless limit restricts the cubic interaction to be equivalent to the on-shell (minimal coupling) given in \eqref{mincouponshell}. 

Little group covariance leads to the following two types of on-shell couplings
\begin{equation}\begin{aligned}
    &i \xi_1(
    h_{\mu\nu } \psib_\rho \gamma^\mu\partial^\nu \psi^\rho-
    h_{\mu\nu }\partial^\nu  \psib_\rho \gamma^\mu\psi^\rho)\,,
\\& i \xi_2(
    h_{\mu\nu } \psib^\rho \gamma^\mu\partial_\rho \psi^\nu-
    h_{\mu\nu }\partial_\rho \psib^\nu \gamma^\mu\psi^\rho)\,,
\end{aligned}\label{mincoupa1a2}
\end{equation}
where the coefficients $\xi_1$, $\xi_2$ have mass dimension $-1$ which are assumed to be $\propto 1/M_P$. 
The question we would like to address is whether the condition $\partial^\mu J_\mu=\mathcal{O}(m_{3/2})$  fixes the ratio $\xi_1/\xi_2$. First, \eqref{mincoupa1a2} contributes to the current divergence as
\begin{equation}\partial^\mu J_\mu\supset
    2i  \xi_1 \partial_\rho h_{\mu\nu}\gamma^\mu \partial^\nu \psi^\rho+i \xi_2  (\partial_\rho h_{\mu\nu}\gamma^\mu \partial^\rho  \psi^\nu +\partial_\rho h_{\mu\nu}\gamma^\mu \partial^\nu  \psi^\rho )+\mathcal{O}(m_{3/2})\,.
\label{mincoupdJ}\end{equation}

An off-shell coupling can be written as the linear combination in \eqref{mincoupa1a2}, plus other terms that vanish on-shell. For the latter, we enumerate all independent possibilities and their contributions to the divergence of the current. Notice that in the definition of the current $J_\mu \equiv \delta V/\delta \psib^\mu$ ($V$ being the three-point vertex), the graviton is on-shell, thus the couplings that vanish on-shell due to the graviton equation of motion will not contribute to $J_\mu$, and consequently will not be listed. One such example is $\partial^\mu h_{\mu\nu} \psib^\rho \gamma^\nu\psi_\rho$ and the vertices related by integration by parts.  

For dimension $4$ interactions, either they need a dimension $5$ interaction to make an on-shell vanishing vertex (see below), or they do not contribute to the current $J_\mu$ because they include the trace $h$. The only exception is
\begin{equation}
    i\xi_3h_{\mu\nu}(\psib^\mu \gamma^\nu \gamma^\rho\psi_\rho-\psib_\rho\gamma^\rho\gamma^\nu  \psi^\mu )\,,
\end{equation}
where $\xi_3$ has mass dimension $0$. This interaction leads to the divergence 
\begin{equation}
      \partial^\mu J_\mu \supset -i \xi_3\partial_\mu h_{\rho\nu}\gamma^\mu\gamma^\nu \psi^\rho -i \xi_3 h_{\rho\mu}\partial^\mu\psi^\rho +\mathcal{O}(m_{3/2})\,. \label{dim4J}
\end{equation}
Since no other contributions from dimension 4 operators can cancel the above terms, 
\eqref{dim4J} can be rendered $\mathcal{O}(m_{3/2})$ only if $\xi_3\propto \kappa  m_{3/2}  $.

For dimension $5$ operators, with one gamma matrix, we can construct the following couplings:
\begin{enumerate}
    \item $i \xi_4 (h_{\mu\nu } \psib_\rho \gamma^\rho \partial^\nu \psi^\mu-h_{\mu\nu }\partial^\nu \psib^\mu  \gamma^\rho\psi _\rho )$
    \item $i \xi_5 (h_{\mu\nu } \psib^\mu \gamma^\nu \partial^\rho \psi_\rho-h_{\mu\nu }\partial^\rho \psib_\rho  \gamma^\nu\psi^\mu )$
    \item $i \xi_6 (h_{\mu\nu } \psib^\mu \gamma^\rho \partial_\rho \psi^\nu-h_{\mu\nu }\partial_\rho \psib^\nu  \gamma^\rho\psi^\mu )-2 \xi_6 m_{3/2} h_{\mu\nu}\psib^\mu \psi^\nu$
\end{enumerate}
They are all vanishing when $\psi_\mu,\psib_\mu$ are on-shell. Their contributions to the current divergence are
\begin{equation}
    \partial^\mu J_\mu \supset i (\xi_4+\xi_6) \partial_\rho h_{\mu\nu}\gamma^\rho \partial^\mu \psi^\nu+\mathcal{O}(m_{3/2})\,.
\label{gdJ}
\end{equation}
Another possible tensor structure  is 
$h_{\cdot \cdot }\psib_\cdot  \gamma_\cdot\gamma_\cdot\gamma_\cdot\partial_\cdot \psi_\cdot+\text{h.c.}$, where the dots are unspecified indices. Structures with more gamma matrices can always be reduced to a combination of one or three gamma matrices using the Clifford algebra. We find the following three independent couplings:
\begin{enumerate}
    \item $i \xi_7 h_{\mu\alpha}(\psib^\alpha \gamma^\mu\gamma^\nu \gamma^\rho\partial_\nu \psi_\rho-\partial_\nu \psib_\rho
    \gamma^\rho\gamma^\nu \gamma^\mu \psi^\alpha)$\item 
     $i \xi_8 h_{\mu\alpha}(\psib_\nu \gamma^\mu\gamma^\nu \gamma^\rho\partial^\alpha \psi_\rho-\partial^\alpha \psib_\rho
    \gamma^\rho\gamma^\nu \gamma^\mu \psi_\nu)$
    \item 
     $i \xi_9 h_{\mu\alpha}(\psib_\nu \gamma^\nu\gamma^\mu \gamma^\rho\partial_\rho \psi^\alpha-\partial_\rho \psib^\alpha
    \gamma^\rho\gamma^\mu \gamma^\nu \psi_\nu)$
\end{enumerate}
Other orderings of the gamma matrices are related by the Clifford algebra to the above couplings and those from $\xi_4 $ to  $\xi_6$. They contribute to the current divergence as
\begin{equation}
       \partial^\mu J_\mu \supset 2i \xi_7 \partial_\mu h_{\rho\nu}\gamma^\nu \partial^\mu \psi^\rho + 2i \xi_8 \partial_\rho h_{\mu\nu}\gamma^\rho \partial^\mu \psi^\nu +\mathcal{O}(m_{3/2})\,.
       \label{gggdJ}
\end{equation}
Cancelling \eqref{mincoupdJ}, \eqref{gdJ} and \eqref{gggdJ} leads to the relations
\begin{equation}
      \xi_2=-2\xi_7\,,\qquad \xi_4+\xi_6=-2\xi_8\,,\qquad \xi_2=-2\xi_1\,.
      \label{eq:xieqns}
\end{equation}
The first two relations in \eqref{eq:xieqns} constrain the couplings that are vanishing on-shell, whereas the last relation  pins down the ratio 
$\xi_1/\xi_2=-1/2$. Using \eqref{mincoupa1a2} this gives rise to the minimal coupling \eqref{mincouponshell}.

\section{All-line transverse shift}
\label{app:recursion}  

In this appendix, we review the all-line transverse momentum shift introduced in Refs~\cite{Ema:2024rss,Ema:2024vww} for the case of spin $\leq1$ particles. The ALT shift is defined by
\begin{equation}
    \hat{p}_i=p_i + z r_i\,,
    \label{alllineshift}
\end{equation}
where $r_i=c_i\varepsilon_i^+ (c_i\varepsilon_i^-)$ for an external particle $i$ with positive (negative) helicity, and $c_i$ are constants. 
Unlike the massless BCFW shift, the shift \eqref{alllineshift} applies to \textit{all} external particles. 
The particular feature of this shift is that the momentum is deformed by the transverse polarization vector, which   will be crucial for connecting the Ward identity to constructibility.

The ALT  shift can also be expressed in the helicity basis. 
Recall that the transverse polarization vectors from \eqref{eq:masslesspolvect} are given by  
\begin{equation}
    \varepsilon^+_{i,\alpha\dotalpha}=\frac{\sqrt{2}}{m} \eta_{i,\alpha}\Tilde{\lambda}_{i,\dotalpha},\qquad 
   \varepsilon^-_{i,\alpha\dotalpha}=\frac{\sqrt{2}}{m} \lambda_{i,\alpha}\Tilde{\eta}_{i,\dotalpha}\,,
   \label{transversepols}
   \end{equation}
and the momentum $p_{i,\alpha\dotalpha}=\lambda_{i,\alpha}\Tilde{\lambda}_{i,\dotalpha}-\eta_{i,\alpha}\Tilde{\eta}_{i,\dotalpha}$. The shift \eqref{alllineshift} then amounts to 
\begin{equation}
    \lambda_{i\alpha}\rightarrow   \lambda_{i\alpha}+ w \eta_{i\alpha},\quad \Tilde{\eta}_{i,\dotalpha}\rightarrow \Tilde{\eta}_{i,\dotalpha}+\tilde{w} \Tilde{\lambda}_{i,\dotalpha},\label{positiveshift}
\end{equation}
if particle $i$ has positive helicity, and
\begin{equation}
\eta_{i\alpha}\rightarrow   \eta_{i\alpha}+\tilde{w}  \lambda_{i\alpha},\quad \Tilde{\lambda}_{i,\dotalpha}\rightarrow \Tilde{\lambda}_{i,\dotalpha}+w \Tilde{\eta}_{i,\dotalpha},\label{negativeshift}
\end{equation}
if particle $i$ has negative helicity. The constants $w,\tilde{w}$ are identified via the relation $w-\tilde{w}=\sqrt{2}zc_i/m$ in order to match \eqref{alllineshift}, with the requirement that $w\neq \tilde{w}$ because otherwise the momentum is not shifted\footnote{For transverse polarizations, the choice of $w$, $\tilde{w}$ is free  provided $w\neq {\tilde w}$. However, this choice will be relevant for the longitudinal state of  spin $1$.}.  The on-shell condition is satisfied, namely
\begin{equation}
    \hat{p}_i^2={p}_i^2+2z p_i\cdot r_i +z^2 r_i^2=p_i^2=m_i^2\,,
\end{equation}
since $r_i\propto \varepsilon^{(\pm)}_i$ are light-like ($r_i^2=0$) and transverse ($p_i\cdot r_i=0$).
The momentum conservation  is fulfilled as well
\begin{equation}
    \sum_i\hat{p}_i=\sum_i p_i+ z \sum_i c_i \varepsilon_i^{(\pm)}=0\,,
\end{equation}
where the equation $\sum_i c_i \varepsilon_i^{(\pm)}=0$ must be imposed. 
Since the temporal component of the transverse polarization is zero, we have  three equations for four unknown variables, thus we are left with one degree of  freedom that is parameterized by the complex variable $z$. 

To prove that the boundary term vanishes ($B_\infty=0$) under the ALT shift, one has to identify all the $z$-dependence in the amplitude. We start by reviewing the spin 1/2, 1 cases and the spin 3/2 will be treated in section~\ref{sec:alt32}. 

If the external state has spin $1/2$, its wave function (defined in \eqref{uvlam}) will appear as $\lambda,\Tilde{\eta}$ for helicity $-1/2$, and $\eta,\Tilde{\lambda}$ for helicity $+1/2$. Following the prescription, we shift $\eta,\Tilde{\lambda}$ for the former case, and shift  $\lambda,\Tilde{\eta}$   for the latter, so that the wave function has no $z$-dependence. For spin $1$,  inspecting  \eqref{transversepols}, we can shift   $\lambda,\Tilde{\eta}$ for helicity $+1$ and $\eta,\Tilde{\lambda}$ for helicity $-1$ so that the external polarization vector is $z$-independent. The special case is helicity zero with polarization vector
\begin{equation}
\varepsilon_{\alpha\dotalpha}^0=\frac{1}{m}
(\lambda_\alpha\Tilde{\lambda}_{\dotalpha}+\eta_\alpha\Tilde{\eta}_{\dotalpha})\,.
\end{equation}
We can choose to shift either by \eqref{positiveshift} or \eqref{negativeshift}, with the choice $w=-\tilde{w}$, so that $\varepsilon^0$ also remains unchanged. In summary, the ALT shift does not modify the external wave functions or polarization vectors for spin 1/2 and 1. 

Next, to discuss the large-$z$ behavior, we follow the dimensional analysis argument in  \cite{Cheung:2015cba}, by decomposing a generic $n$-point amplitude into
\begin{equation}
    A_n = \left(\sum_{\mathrm{diagrams}}g \times F\right) \times \prod_{\mathrm{vectors}} \varepsilon \times \prod_{\mathrm{fermions}}u\,,
\end{equation}
where $g$ represents the coupling and $F$ is the contribution from Feynman rules with external polarization vectors and wave functions stripped off. The contribution $F$ is typically comprised of momentum insertions,  the propagator, polarization sum of the internal particle, etc. The contributions from $\varepsilon$, $u$ represent the external polarization vectors and fermion wave functions, respectively. In particular, each vector-spinor $\psi^\mu$ will contribute with one $\varepsilon$ and one $u$. 

Analyzing the mass dimensions, we have
\begin{equation}
    4-n = [g] + [N] - [D] + \frac{1}{2}N_\chi\,,
\end{equation}
where $N_\chi$ is the number of fermions and $N, D$ are respectively, the numerator and denominator of $F$. Since each momentum acquires a linear dependence on $z$ (at large $z$) under the all-line shift in \eqref{alllineshift}, the propagator thus behaves as $D\sim z^{-[D]}$ at large $z$.  The numerator may contain momenta from derivative couplings that are also shifted, so it grows at large $z$ as $N \sim z^{\gamma_N} $ with $\gamma_N \leq [N]$.  In total, the large-$z$ behavior of the amplitude $A_n \sim z^{\gamma+\rho_n}$ where 
 \begin{equation} 
	\gamma = \gamma_N - [D] \leq [N] - [D] = 4-n - [g] - \frac{N_\chi}{2}\,,
 \label{naivegamma}
 \end{equation}
and $\rho_n$ encodes the $z$-dependence in $\varepsilon$ and $u$. For  spin 1/2 and spin 1, we have seen previously that $\rho_n=0$, independently of the helicity. Therefore, it is easy to see that if $\gamma<0$, the boundary term vanishes, so the amplitude is \textit{constructible}. Restricting to a renormalizable theory, the coupling constants have the dimension $[g]\geq0$, then any $n$-point amplitude with $n\geq5$ is constructible by the ALT shift. The four-point amplitude involving at least one fermion is constructible as well.

The special case is the four-point amplitude with massive spin-1 particles and $[g]=0$. From \eqref{naivegamma} one may naively conclude that $\gamma\leq0$ hence the amplitude is not necessarily constructible. In fact, this upper bound can be improved if one \textit{imposes} the Ward identity. Suppose there is one external transverse state that is labelled ``1'', then the amplitude can be decomposed into
\begin{equation}   
A_4=\varepsilon^{\mu(\pm)}_1F_\mu(p_1,p_2,p_3,p_4)\,.
\end{equation}
The leading $z$-dependence corresponds to replacing all momenta by the shifted ones, namely
\begin{equation}
   \lim_{z\rightarrow\infty} \hat{A}_4 = z^{0} \varepsilon^{\mu(\pm)}_1 F_\mu(r_1,r_2,r_3,r_4)\,,
\end{equation}
where now $F_\mu(r_1,r_2,r_3,r_4)$ is the stripped amplitude with massless dynamics because $r_i^2=0$. In addition, under the ALT shift, the shifted momenta are proportional to the transverse polarization vector $r_1^\mu\propto \varepsilon^{\mu(\pm)}_1$, therefore, 
\begin{equation}
   \lim_{z\rightarrow\infty} \hat{A}_4 \propto z^{0} r_1^\mu F_\mu(r_1,r_2,r_3,r_4)=0\,.
\end{equation}
The last equality holds thanks to Ward identity, which then makes the amplitude constructible.

Given that $B_\infty=0$, one can now solve the equation $\hat{p}_I^2=m_I^2$ and use Eq.~\eqref{four-point-construction} to construct four-point amplitudes. The pole condition   is quadratic in $z$, hence has two solutions  $z_\pm$. The product of them satisfies
\begin{equation}
    \begin{aligned}
	z_{ij}^+ z_{ij}^- = \frac{p_{ij}^2 - m_I^2}{2r_i \cdot r_j}\,.
\end{aligned}\label{zpluszminus}
\end{equation}
Inserting the solutions in the residue formula, the four-point amplitude becomes
\begin{equation}
\begin{aligned}
A_4^{(\lambda_1 \lambda_2 \lambda_3\lambda_4)}
	&= \sum_{I=\text{\tiny internal particles}} \sum_{(i,j) = (1,2), (1,3), (1,4)}\frac{1}{p_{ij}^2-m_I^2}
 \\	&\times
	\sum_{\lambda}
	\left[
	\frac{z^+_{ij}}{z^+_{ij} - z^-_{ij}} \hat{A}^{(\lambda_i \lambda_j \lambda)}(z^-_{ij}) 
	\times \hat{A}^{(-\lambda \lambda_k \lambda_l)}(z^-_{ij})
	- \frac{z^-_{ij}}{z^+_{ij} - z^-_{ij}} \hat{A}^{(\lambda_i \lambda_j \lambda)}(z^+_{ij}) 
	\times \hat{A}^{(-\lambda \lambda_k \lambda_l)}(z^+_{ij})
	\right],
\end{aligned}
\label{fact-and-cont}
\end{equation} 
where $l,k\neq i,j$, and all $(s,t,u)$ channels are accounted for (the formula can be easily adapted to smaller number of channels). In the second sum, when the different channels are related by the exchange of   Majorana fermions, there will be an extra relative minus sign. Note that any linear dependence on $z$ in $\hat{A}_3(z)\hat{A}_3(z)$ cancels   according to the above formula. 
For applications to QED and electroweak amplitudes, see \cite{Ema:2024rss,Ema:2024vww}.

\newpage
\bibliographystyle{utphys}
\bibliography{refs}
\end{document}